\PassOptionsToPackage{desactivate}{linenoaa}
\documentclass[longauth]{aa} 

\usepackage{graphicx}   
\usepackage{rotating}   
\usepackage{lscape}     

\usepackage{float}

\usepackage[dvipsnames]{xcolor}

\usepackage{txfonts}
\usepackage{placeins}
\usepackage[normalem]{ulem}

\def\gs{\mathrel{\raise0.35ex\hbox{$\scriptstyle >$}\kern-0.6em \lower0.40ex\hbox{{$\scriptstyle \sim$}}}}
\def\ls{\mathrel{\raise0.35ex\hbox{$\scriptstyle <$}\kern-0.6em \lower0.40ex\hbox{{$\scriptstyle \sim$}}}}

\newcommand{\orcid}[1]{\includegraphics[scale=0.06]{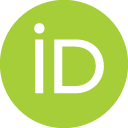} \href{https://orcid.org/#1}{#1}}

\usepackage{hyperref}
\usepackage{multirow}

\begin{document} 

\title{MIDIS: The identification of deep MIRI-red sources as candidates for extreme Balmer-break and line emitting galaxies at high-$z$}
   
\titlerunning{MIRI-red sources at 5.6\,$\mu$m }
\author{I. Jermann \inst{1,2,3} %
        \and G. Brammer\inst{2,3}
        \and S. Gillman\inst{1,2}
        \and T. R. Greve\inst{1,2}
        \and L. A. Boogaard\inst{4}
        \and J. Melinder\inst{5}
        \and R. A. Meyer\inst{6}
        \and P. G. P\'erez-Gonz\'alez \inst{7}
        \and P. Rinaldi\inst{8}
        \and L. Colina\inst{7}
        \and G. \"Ostlin\inst{5}
        \and G. Wright\inst{9}
        \and J. \'Alvarez-M\'arquez\inst{7}
        \and A. Bik\inst{5}
        \and K. I. Caputi\inst{10}
        \and L. Costantin\inst{7}
        \and A. Crespo G\'omez\inst{8}
        \and J. Hjorth\inst{11}
        \and E. Iani\inst{12}
        \and S. Kendrew\inst{13}
        \and A. Labiano\inst{7,14}
        \and D. Langeroodi\inst{11}
        \and F. Peissker\inst{15}
        \and C. Prieto-Jiménez\inst{7,16}
        \and J. P. Pye\inst{17}
        \and T. V. Tikkanen\inst{17}
        \and F. Walter\inst{18}
        \and P. van der Werf\inst{4}
        \and T. Henning\inst{18}
        \and M. Shuntov\inst{2,3}
          }

   \institute{
        Cosmic Dawn Center (DAWN), Denmark\\
        \email{irisj@space.dtu.dk}
        \and
        DTU-Space, Technical University of Denmark, Elektrovej 327, DK-2800 Kgs. Lyngby, Denmark
        \and 
        Niels Bohr Institute, University of Copenhagen, Jagtvej 128, DK-2200, Copenhagen, Denmark
        \and 
        Leiden Observatory, Leiden University, PO Box 9513, NL-2300 RA Leiden, The Netherlands
        \and 
        Department of Astronomy, Oskar Klein Centre, Stockholm University, AlbaNova University Center, 10691 Stockholm, Sweden 
        \and 
        Department of Astronomy, University of Geneva, Chemin Pegasi 51, 1290 Versoix, Switzerland
        \and  
        Centro de Astrobiología (CAB), CSIC-INTA, Ctra. de Ajalvir km 4, Torrejón de Ardoz 28850, Madrid, Spain
        \and
        Space Telescope Science Institute, 3700 San Martin Drive, Baltimore, MD 21218, USA
        \and 
        UK Astronomy Technology Centre, Royal Observatory Edinburgh, Blackford Hill, Edinburgh EH9 3HJ, UK
        \and
        Kapteyn Astronomical Institute, University of Groningen, PO Box 800, 9700 AV Groningen, The Netherlands
        \and
        DARK, Niels Bohr Institute, University of Copenhagen, Jagtvej 155A, 2200 Copenhagen, Denmark    
        \and 
        Institute of Science and Technology Austria (ISTA), Am Campus 1, 3400 Klosterneuburg, Austria
        \and 
        European Space Agency, Space Telescope Science Institute, Baltimore, MD, USA
        \and
        Telespazio UK for the European Space Agency (ESA), ESAC,
        Camino Bajo del Castillo s/n, 28692 Villanueva de la Cañada, Spain
        \and
        Physikalisches Institut der Universität zu Köln, Zülpicher Str. 77, 50937 Köln, Germany
        \and
        Departamento de Física de la Tierra y Astrofísica, Facultad de Ciencias Físicas, Universidad Complutense de Madrid, E-28040 Madrid, Spain    
        \and
        School of Physics \& Astronomy, Space Park Leicester, University of Leicester, 92 Corporation Road, Leicester LE4 5SP, UK 
        \and 
        Max-Planck-Institut für Astronomie, Königstuhl 17, 69117 Heidelberg, Germany}

   \date{Received MMM dd, yyyy; accepted MMM dd, yyyy}

  \abstract
  { We investigate the detection and nature of 5.6\,$\mu$m MIRI-red sources in the MIRI Deep Imaging Survey (MIDIS), covering 2.4 arcmin$^2$ in the Hubble Ultra Deep Field. MIDIS is the deepest \textit{JWST}/MIRI survey to date, probing the faintest limits and enabling the study of rare galaxy populations at high redshift. We define MIRI-red sources as detected at $5\sigma$ significance in MIRI/F560W with red colors: $m_{\rm F444W}-m_{\rm F560W} \geq 0.5$. We characterize the MIDIS background to assess the effective depth and noise properties of faint MIRI detections, finding that pipeline uncertainties lead to a systematic overestimation of the signal-to-noise ratio by a factor of $\sim\,3$. Using an empirical methodology, we estimate the purity and completeness of MIRI detections and find that a $5\sigma$ detection at 28.75 mag has a purity of 92\% and a completeness of 54\%. We identify seven MIRI-red galaxy candidates, including an F115W dropout consistent with a high-redshift galaxy candidate. We explore possible physical origins for the MIRI-red population, including active galactic nuclei, dust-obscured galaxies, extreme emission-line galaxies, evolved stellar populations, and Little Red Dots (LRDs). Given the proximity of the F444W and F560W filters and the depth of MIDIS, we find that MIRI-red galaxy candidates are consistent with either emission-line galaxies with EW$_0({\rm H\alpha})\geq 750$ Å or EW$_0{(\rm H\beta+[OIII])}\geq 600$ Å, or high-redshift Balmer breaks of at least 1.6. We additionally discuss the extreme case of a MIRI-red galaxy candidates undetected in F444W, a potential MIRI-only source, for which we derive extreme line strengths of EW$_0({\rm H\alpha})\sim6000$ Å and EW$_0({\rm H\beta+[OIII]}) \sim 4000$ Å, or high-$z$ LRD analogs exhibiting Balmer breaks of 6.3. Finally, we find fewer MIRI-red detections than expected from extrapolations of the H$\alpha$ or H$\beta$+[OIII] emitter line luminosity functions, consistent with previous deep searches, while the absence of $z>10$ LRD candidates is consistent with theoretical expectations for the MIDIS survey volume. }


\maketitle

\section{Introduction}
\label{sec:Introduction}

Probing the universe deep beyond the cosmic reionization with an unprecedented sensitivity and resolution is one of the greatest achievements of the \textit{James Webb Space Telescope (JWST}, \cite{Gardner2006_JWST,Gardner2023_jwst}). The near-infrared camera (NIRCam, \cite{Rieke2023_NIRCam_citation}) and the near-infrared spectrograph (NIRSpec, \cite{Boker2023_NIRSpec_citation}) have confirmed hundreds of galaxies in the cosmic reionization \citep{Wang2023_uncover,Carniani2024,Harikane2024_spectroUVLF,Roberts-Borsani2024_lines,DEugenio2025_jades_data}. A key trade-off in extragalactic surveys lies between wide, shallow programs that sample the bright end of the luminosity function, and ultra-deep pencil-beam observations that reach intrinsically faint systems, the sub-$L_\star$ galaxies. Deep surveys have therefore played a crucial role in pushing the observable frontier of galaxy formation and understanding the bulk of the galaxy populations. Remaining open questions are how much the faint systems contribute to the cosmic reionization and to the total cosmic star formation rate density (CSFRD) at high-$z$, e.g., \cite{Chemerynska2025_faint_gal} suggest that $M_{\rm UV}<-19$ would contribute up to 50\% to the total CSFRD at $z>9$ and \cite{Matthee2024,Madau2024_reioni} question the role of faint and high-$z$ active galactic nuclei (AGN) in the cosmic reionization.

A landmark survey for probing the sub-$L\star$ galaxy populations is the Hubble Ultra Deep Field (HUDF; \citet{Beckwith2006HUDF}). The HUDF is a Cycle 5 \textit{Hubble Space Telescope (HST)} program that provided ultra-deep observations in four \textit{HST}/ACS bands \citep{Williams1996HDF}, resulting in a unique field with exceptionally deep and multi-wavelength coverage from many ground- and space-based telescopes. The \textit{HST} observations that cover the far-UV \citep{Voyer2009_HUDF_farUV}, near-UV \citep{Teplitz2013_HUDF_nearUV}, and near-IR \citep{Oesch2010_HUDF_nearIR,Bouwens2011_HUDF_nearIR,Ellis2013_HUDF_nearIR,Koekemoer2013_HUDF_nearIR} are collectively known as the eXtreme Deep Field (XDF; \citet{Illingworth2013_XDF,Koekemoer2013_HUDF_nearIR}), reaching 5$\sigma$ depths of $\sim29$–$30$ AB magnitudes in optical photometric bands (e.g., >30 in ACS/F606W and $\sim$29.5 in WFC3/F160W). With the advent of the \textit{JWST}, the HUDF was naturally observed with deep early NIRCam and NIRSpec programs (FRESCO \citet{Oesch2023_FRESCO}, JADES \citet{Eisenstein2023_JADES,Rieke2023_JADES_nircam,Bunker2024_JADES_nirspec_z11REF,DEugenio2025_jades_data}, and JEMS \citet{Williams2023_JEMS}), adding deep photometric and spectroscopic coverage up to 5\,$\mu$m in this particular region of the sky.


High-redshift galaxies have been identified using the dropout technique, most notably through the Lyman-break selection, which exploits the strong absorption blueward of Ly$\alpha$ by neutral hydrogen in the intergalactic medium \citep{Steidel1996,Bromm2011_review,Dayal2018,Bouwens2011_HUDF_nearIR, Ellis2013_HUDF_nearIR, Coe2013}. With \textit{JWST}, this approach has been significantly extended to higher redshifts \citep{Treu2022,Finkelstein2023,Eisenstein2023_JADES,Bagley2024,Bezanson2024,Donnan2024}. Broadband spectral breaks can also arise from strong nebular emission lines boosting individual filters \citep{RobertsBorsani2020,Laporte2021,Laporte2023}, or exploiting the wide wavelength coverage of the \textit{JWST}, it is possible to combine the search for the Lyman and Balmer/4000\,\AA\ breaks in young and rapidly evolving stellar populations \citep{Henry2008_DoubleBreakSpitzer,Desprez2024}. In particular, extreme emission-line galaxies can exhibit very red near- to mid-infrared colors when rest-frame optical lines such as H$\alpha$ or H$\beta$+[OIII] enter a single filter, mimicking classical dropout signatures, e.g., \citet{Naidu2022,Zavala2023,vanMierlo2024}. In the standard hierarchical $\Lambda$CDM structure formation model, primeval objects such as Population III stars (Pop III) and direct collapse black holes (DCBHs) show strong hydrogen recombination lines that in principle can lead to a photometric dropout \citep{Bromm2004,Schneider2006,Zackrisson2011_JWST,Nakajima2022_popIII_DBHC}.

Photometric dropouts have enabled the discovery of new galaxy populations with \textit{JWST}: the Little Red Dots (LRDs). LRDs are identified for their red compact morphology, broad H$\alpha$ emission line, and a characteristic V-shape of the spectral continuum \citep{Ubler2023_LRD,Harikane2023_LRD,Kokorev2023_LRD,Furtak2024_LRD,Greene2024_LRD,Matthee2024,Kocevski2025_LRDstuff}. A physical interpretation for LRDs is a host galaxy dominated by the emission of an early supermassive black hole, a "black hole star", in the process of forming an atmosphere from turbulent and very dense gas \citep{Naidu2025BHstar,InayoshiMaiolino2025,deGraaff2025_BHstar,Rusakov2026_Nature,Caputi2026}. The wavelength coverage of MIRI opens the possibility of identifying sources that are faint or undetected at $\lambda<5\,\mu$m but become detectable in the mid-infrared, raising whether a population of extreme, line-dominated, dust-obscured, or accreting systems may be missed by previous near-infrared instrument alone such as \textit{Spitzer} \citep{PerezGonzalez2024_cerberus,Iani2025_Virgil,Bing2025arXiv251108672B}.

As a Guaranteed Time Observation Cycle 1 program, the European MIRI consortium obtain a deep image in the HUDF at 5.6\,$\mu$m using the imager mode (MIRIM, \citet{Bouchet2015_jwst_mirim}, \citet{Dicken2024_jwst_mirim}) of MIRI, probing in the Mid-Infrared at > 28 mag. As comparison, for NIRCam need $\sim 2$ hours to reach 28 mag in the reddest band F444W and MIRI is $\sim 50$ hours to reach the same depth in the bluest band at F560W \citep{Ostlin2025}. In this work, we push and explore the limits of the \textit{JWST} observing capabilities by searching for 5\,$\mu$m photometric dropout. We search for 5.6\,$\mu$m MIRI-red galaxy candidates in the HUDF and investigate their nature. We predict the expected number assuming different galaxy populations and compare to our survey number counts and candidate sample.

This paper is structured as follows. In Sect.~\ref{sec:Data}, we present the MIri Deep Imaging Survey (MIDIS; \cite{Ostlin2025}) observations and the ancillary multi-wavelength data used in our analysis. In Sect.~\ref{sec:Method}, we present an extensive study of the MIDIS background noise (Sect.~\ref{sec:MIRInoise}), we describe our catalog for >23.5 mag MIRI sources (Sect.~\ref{sec:MIDISdeepCatalog}), and we measure MIDIS purity and completeness (Sect.~\ref{sec:MIDISdeepPurityCompleteness}). In Sect.~\ref{sec:Results}, we report the MIRI-red galaxy candidates in MIDIS (\ref{sec:ResultsMIRICandidates}) and characterize their potential nature in the context of \textit{JWST} spectroscopic observations and spectral energy distribution (SED) models (\ref{sec:ResultsMIRIonlyCharachterization}). We discuss the nature of MIRI-red sources in Sect.~\ref{sec:OnTheNature} and compare the MIDIS number counts to emission line luminosity functions and LRD expected abundance. Sect.~\ref{sec:Summary} consists of a summary of our work and findings.

We adopt the following cosmology: flat $\Lambda$CDM cosmology with H$_0$ = 70\,km\,s$^{-1}$ Mpc$^{-1}$, $\Omega_0$ = 0.3, consistent with the most recent measurements from Planck \citep{Planck2016}. The magnitudes are given in the AB system \citep{Oke1983}.

\section{Data}
\label{sec:Data}
 
To identify MIRI-red galaxy candidates in the MIDIS survey of the HUDF, we first define the search region within the survey of MIRI-red candidates (Sect.~\ref{sec:MIDIS}). We then compile the rich multi-wavelength ancillary observations of the field (Sect.~\ref{sec:OtherData}) and measure their depths (Sect.~\ref{sec:Depths}).

\subsection{The MIRI Deep Imaging Survey (MIDIS)}
\label{sec:MIDIS}

\begin{figure}[h] 
   \centering
  \includegraphics[width=\hsize]{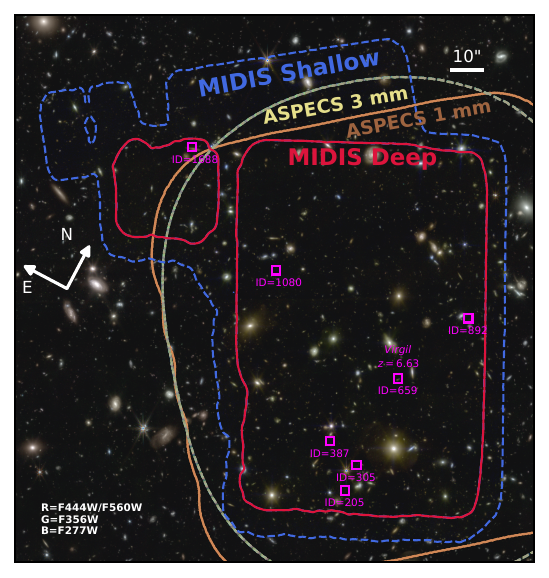}
    \caption{RGB image of the MIDIS field constructed with \textsc{trilogy} \citep{Coe2012_Trilogy}, combining F560W within the MIDIS footprint and F444W outside of it. Red solid and blue dashed contours indicate MIDIS Deep and Shallow regions, respectively. Yellow and brown contours show Band 3 (3 mm) and Band 6 (1 mm) ALMA observations \citep{ASPECS1_Walter2016,GonzalezLopez2019_ALP1}. Magenta squares mark MIRI-red galaxy candidates and their IDs.}
\label{fig:LETTER_MIRIonlyFigure1}
\end{figure}

The observations for MIDIS (Program ID \href{https://www.stsci.edu/jwst/science-execution/program-information?id=1283}{1283}, PIs Hans Ulrik Nørgaard-Nielsen, G.~Östlin), a GTO cycle 1 program from the European MIRI Consortium, were carried out on 2$^\text{nd}$\,--\,6$^\text{th}$, and 20$^\text{th}$ of December 2022 using the F560W filter with of MIRIM \citep{Ostlin2025}. The MIDIS survey consists of $\sim$50-hours at F560W and $\sim$10-hours in F1000W that have been observed in December 2023 and decided to be carried out in that band due to a safe mode event that interrupted and delayed the original observing schedule. MIDIS is the deepest MIRI observations to date consisting of 96 exposures with a total 52.75 hours on source observation in a single MIRI pointing located in a part of the HUDF and of the XDF (see Fig.~\ref{fig:LETTER_MIRIonlyFigure1}), specifically chosen to overlap with the UDF-10 region covered by the MUSE \textit{Hubble} Ultra Deep Field Survey \citep{Bacon2017} and the ALMA SPECtroscopic Survey in the \textit{Hubble} Ultra-Deep Field pilot and large programs \citep{ASPECS1_Walter2016,GonzalezLopez2019_ALP1}, herafter MXDF and ASPECS.

As discussed in \citet{Ostlin2025}, the MIDIS/F560W image (first internal release, v1.4) was reduced using a modified version of the official JWST pipeline\footnote{\url{https://github.com/spacetelescope/jwst}}, version v1.12.3 (CRDS\_CONTEXT\,=\,1137) and additional routines were performed to minimize the effect of cosmic showers, instrument noise, and background variations on the final mosaic \citep{PerezGonzalez2024_MIRIreduction}. The full reduction procedure and an overview of MIDIS observations are presented in \citet{Ostlin2025}. Due to a safe-mode event and the rescheduling of the observation plan, the final MIDIS F560W image consists of exposures with different position angles and exhibits inhomogeneous depth. We divide the remaining image into two distinct areas accordingly to the 35$^{\rm th}$ and 60$^{\rm th}$ percentiles of the RMS (SCI/ERR) pixel distribution: MIDIS Deep and MIDIS Shallow\footnote{The MIDIS Deep footprint is consistent with region A+B in \citet{Ostlin2025} who defined regions according to the observation time. Our depth measurement is consistent with \citet{Ostlin2025} and slight variations are due to subtle variations in the methodologies and the exact sub-regions.}, as illustrated in Fig.~\ref{fig:LETTER_MIRIonlyFigure1}. This paper focuses on the MIDIS Deep region that reaches a 5\,$\sigma$ depth of 28.75 mag, 11.5\,nJy (Sect.~\ref{sec:Depths}, Table~\ref{table:DepthTable}).

As a continuation of the MIDIS survey, the European MIRI Consortium high-redshift team has
been awarded a Cycle 3 GO program (MIDIS-RED; Program
ID \href{https://www.stsci.edu/jwst/science-execution/program-information?id=6511}{6511}, PI: G. Östlin) that has observed $\sim$\,40 hours in F770W in December 2024 and $\sim55$ hours in the MIRI/F1000W band in December 2025. Thus, a newer version (internal release v.2.2.8) of the MIDIS observation is reduced with a JWST pipeline version 1.16.1, CRDS\_CONTEXT = 1303, along with the MIDIS-RED F770W observations. In early 2026, the MIDIS and MIDIS-RED observation in the band F1000W will be reduced together (Östlin et al. in prep.). The newest reductions of the MIDIS and MIDIS-RED images will be made available on MAST, and versions v.2.2.8, used in this work, are already accessible on Zenodo \citep{MIDIS2025_Zenodo}\footnote{\url{ https://doi.org/10.5281/zenodo.15624625}}.

In this work, we adopt a pixel scale of 0\farcs{04} (2.75 $\times$ smaller than  MIRI native of 0\farcs{11}) motivated to match the pixel scale of the NIRCam images in the same field and ensure a consistent analysis. The analysis of MIRI-red sources in MIDIS/F560W is done on the second version (v.2.2.8) of the image. We address differences in noise properties (Sect.~\ref{sec:MIRInoise}) on the other pixel scale (0\farcs{06}) and the images from the first internal release (v.1.4) in Appendix \ref{sec:AppendixA_MIDIS_versions}. In Figure~\ref{fig:LETTER_MIRIonlyFigure1}, we show an RGB-color image of MIDIS combining it with NIRCam observations of the same field (Sect.~\ref{sec:OtherData}). We overlay the footprints of the MIDIS Deep, MIDIS Shallow, and the deep ALMA observations from ASPECS. The magenta squares indicate the location in the field of the MIRI-red galaxy candidates (Sect.\ref{sec:ResultsMIRICandidates}). For the deep MIDIS-red F770W and F1000W observation, we use the first internal release.

\subsection{Ancillary Multi Wavelength Observations and spectroscopic redshift sources}
\label{sec:OtherData}

In order to search for MIRI-red galaxy candidates, we first collate the imaging multi-wavelength data to MIDIS Deep, \textit{HST}, \textit{JWST}, and ALMA. The rich coverage in the HUDF covers the optical (0.435-0.850\,$\mu$m) with \textit{HST}/ACS (F435W, F606W, F775, F814W, and F850LP), in the near-infrared (1.05-4.44\,$\mu$m) with \textit{HST}/WFC3 (F105W, F125W, F140W, and F160Wr), and \textit{JWST}/NIRCam (wide bands: F090W, F115W, F150W, F200W, F277W, F356W, and F444W, medium bands: F182M, F210M, F410M, F430M, F460M). We obtain the \textit{HST} and \textit{JWST} images from \href{https://dawn-cph.github.io/dja/blog/2023/07/18/image-data-products/}{the DAWN JWST Archive (DJA)}. DJA is a repository of all public \textit{JWST} observations reduced with \textsc{grizli}~\citep{grizli_Zenodo}. The ALMA observations also provide mm coverage (1-3.6\,mm), specifically in Bands 3 (2.6–3.6\,mm) and 6 (1.1–1.4\,mm). We compile the ASPECS galaxies from the catalogs provided by \citet{ASPECS1_Walter2016,Decarli2019_ASPECS,Boogaard2023}. 

We then collect the spectroscopic observation in the MIDIS field of view from MUSE in the optical (0.475-0.930\,$\mu$m) and in the near-IR with NIRSpec/PRISM (0.3-5.3\,$\mu$m) in order to understand the red galaxy population in MIDIS and further investigate the nature of MIRI-only sources. We utilize the MXDF catalog from \citet{Bacon2022Cat}, that was designed to match ASPECS \cite{Bacon2017}\footnote{\url{https://aspecs.info/data/}}. We use the NIRSpec/PRISM catalog from the DJA reduced with \textsc{MSAExp} version 4 \citep{Brammer2023_MSAexp_zenodo,DeGraaff2024_msaexp,Heintz2025_msaexp,Pollock2025}.

We collect 134 NIRSpec/PRISM galaxies with grade 3 $z_{\rm spec}$ that have a match with a NIRCam detection within 0.25" and select 162 galaxies with a confidence estimate for the redshift based on a single or multiple feature (grade$\ge3$ in the DJA catalogue), that match a NIRCam detection in REG 3 within 0.25" from MUSE. 20 of the MUSE sources have a NIRSpec redshift, for which we keep the NIRSpec value\footnote{One redshift disagrees: \texttt{gds-udeep-v4\_prism-clear\_3215\_209051.spec.fits} at $z_{\rm spec}=4.772$ and ID=7396 at $z_{\rm spec}=3.442$.}.

\subsection{ Depths }
\label{sec:Depths}

For consistency in our analysis of NIRCam and MIRI colors, we adopt the same methodology and aperture sizes to determine the depths in MIRI and in the NIRCam images within the footprint of MIDIS Deep. We measure the aperture corrected depths from the standard deviation of fluxes in apertures of diameter D=0\farcs{4} across the background. We report all our measured depths in D=0\farcs{4} apertures in Table~\ref{table:DepthTable}. 

A difference in the MIRI depth with \citet{Ostlin2025} (comparing the first internal release, Appendix \ref{sec:AppendixA_MIDIS_versions}) is due to a different choice in aperture size (D\,=\,0\farcs{45} in their analysis) and from methodological choices. Differences between our NIRCam depth measurements and those reported by other groups can be attributed to the sub-region of MIDIS Deep, the choice in aperture size, aperture correction, and reduction pipeline. Using our methodology with apertures sizes of D=0\farcs{16}, as in \citet{Weibel2024}, and D=0\farcs{2}, as in \citet{Eisenstein2023_JADES,Eisenstein2023b_JADES_DR2}, we measure 1$\sigma$ depth in F444W of 0.49 nJy (non-aperture corrected) and 5$\sigma$ magnitude of 30.43 (also non-aperture-corrected), consistent with the values reported in both studies.

\begin{table}[h!]
\caption{ Aperture corrected (D=0\farcs{4}) depths of MIDIS and NIRCam bands. The NIRCam stack consists of bands F277W+F356W+F444W. F1000W$^\star$ includes the latest F1000W MIDIS-RED observations from December 2025.  }          
\label{table:DepthTable}      
\centering                        
\begin{tabular}{c c c c }        
\hline\hline                

BAND & 5$\sigma$ (mag) & 1$\sigma$ (mag) & 1$\sigma$ [nJy] \\  \hline
 F1000W$^\star$ & 27.18 & 28.92 & 9.79 \\
 F1000W & 26.92 & 28.66 & 12.42  \\ 
 F770W & 28.14 & 29.89 & 4.02  \\ 
 F560W & 28.75 & 30.50 & 2.30  \\ \hline
 F480M & 28.12 & 29.87 & 4.11  \\ 
 F460M & 27.76 & 29.51 & 5.70  \\ 
 F444W & 29.01 & 30.75 & 1.81  \\ 
 F430M & 28.05 & 29.80 & 4.37  \\ 
 F410M & 28.66 & 30.41 & 2.49  \\ 
 F356W & 29.18 & 30.93 & 1.54  \\ 
 F335M & 28.74 & 30.49 & 2.31  \\ 
 F277W & 29.25 & 30.99 & 1.45  \\ 
 F210M & 28.48 & 30.23 & 2.95  \\ 
 F200W & 28.64 & 30.39 & 2.54  \\ 
 F182M & 28.65 & 30.39 & 2.53  \\ 
 F150W & 28.61 & 30.36 & 2.60  \\ 
 F115W & 28.63 & 30.37 & 2.57  \\ 
 F090W & 28.40 & 30.15 & 3.16  \\ 
 NIRCam stack & 29.39 & 31.14 & 1.27  \\  
 
\hline 
      
\end{tabular}
\end{table}

\section{Method}
\label{sec:Method}

In this work, we search for sources detected at 5$\sigma$ at 5.6\,$\mu$m in the MIRI F560W filter. The identification of significant photometric detections in a single photometric band requires a careful analysis. We investigate ways of automating and optimizing the selection of MIRI-red sources, given that the survey design and multi-wavelength coverage, pushes the search toward the fainter end of MIRI detections. In this section, we describe our selection of MIRI-red galaxy candidates, which includes masking of bright sources (Sect.~\ref{sec:StellarMask}), a careful measure of the signal-to-noise (S/N) (Sect.~\ref{sec:MIRInoise}), a catalog for NIRCam detection and a catalog for MIRI detections (Sect.~\ref{sec:MIDISdeepCatalog}), and an estimate of purity and completeness (Sect.~\ref{sec:MIDISdeepPurityCompleteness}), that together lead to our selection criteria for MIRI-red galaxy candidates in MIDIS Deep (Sect.~\ref{sec:SelectionCriteria}).

\subsection{Masking spurious detections near bright sources} 
\label{sec:StellarMask}

The \textit{JWST}/MIRI point-spread function (PSF) has notably extended wings \citep{Rigby2023_forMIRIpsf}, which can lead to spurious detections in the extended emission of bright sources.  Variations in the position angles of the MIDIS survey exposures (due to the scheduling of the observations, see Sect.~\ref{sec:MIDIS}) result in a noisier PSF, with diffraction spikes oriented differently across the image. To account for this, we used the normalized empirical MIDIS PSFs constructed by \citet{Boogaard2023}, publicly available on \href{https://github.com/lboogaard/midis_psf}{GitHub}; further details are discussed therein. 

\begin{figure}[h] 
   \centering
  \includegraphics[width=\hsize]{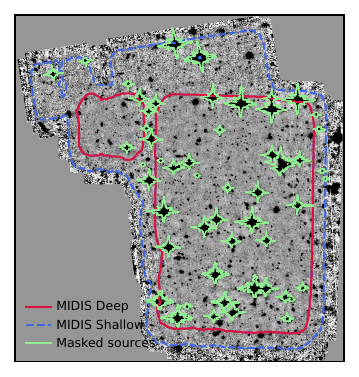}
   \caption{ Illustration of the mask applied to MIDIS/F560W bright sources ($<23.5$,mag). The MIDIS Deep and MIDIS Shallow contour regions are overlaid. After masking the bright galaxies, the remaining area corresponds to 91.3\% of the total MIDIS Deep area. }

\label{fig:MaskFig}
\end{figure}

We constructed a mask for the 61 sources brighter than 23.5 mag, as we expect MIRI-only sources to be fainter given the multi-wavelength coverage and depths. Using a blank image grid matching the MIDIS field, we inserted at the RA and DEC of each bright source the mask created from the respective empirical PSF model with correct position angle. Each mask was scaled proportionally to the \textsc{kron} flux \citep{Kron1980} of its associated source to ensure the full extent of bright sources were covered. The mask covers 9\% of the MIDIS Deep area, and it is homogeneously masking throughout the field. We show the mask in Figure~\ref{fig:MaskFig}. Our search and study of MIRI-red sources in MIDIS Deep were conducted, excluding the masked regions. Note that for example JADES-GS-z10-0 \citep{Hainline2024_z11REF}, a galaxy confirmed with NIRSpec, is blended in F560W with a bright galaxy and is thus masked in our catalog.

\subsection{MIRI noise study}
\label{sec:MIRInoise}

For drizzled observations, it is well known that the noise estimated from variance maps with \textsc{Source-Extractor} \citep{Bertin1996} is underestimated. This is because correlated noise induced by the drizzling process, and other systematics that are not accounted for in the pipeline, such as detector scattering \citep{Gaspar2021_detectorscattering} or charge migration \citep{Argyriou2023_chargemigration}, contribute to the noise and can dominate the faint-source regime. This results in an overestimate of the S/N of faint detections and enhances the significance of faint MIRI-red galaxy candidates by nearly a factor of three (see Sect.\ref{sec:MIRInoiseMIRIonly}).

We develop an empirical method to characterize the noise level in MIDIS from the scatter in the background of the science image (Sect.~\ref{sec:MIRInoiseBackground}). The empirical method is based on the methodologies suggested by \citet{Labbe2003,Gawiser2006,Quadri2007} to characterize noise of faint sources in deep and drizzled observations. These approaches have been adopted in several deep-field and multi-wavelength photometric catalogs, e.g., \citet{Whitaker2011,Skelton2014,Weibel2024,Shuntov2025_cosmosweb}. We study the average error from the variance map and suggest a methodology to calculate a rescaling factor to apply to the variance map when estimating the errors from a software such as \textsc{Source-Extractor}, as an alternative to the empirical method (Sect.~\ref{sec:MIRInoisePipeline}). 

Rescaling the variance map and adopting the empirical method give consistent S/Ns, with the only difference that the empirical method accounts for the correlated noise that does not manifest linearly with aperture size (Sect.~\ref{sec:MIRInoiseMIRIonly}). Throughout our analysis, we adopt the empirical method. 

\subsubsection{MIRI noise from the background }
\label{sec:MIRInoiseBackground}

We estimate the empirical noise, $\sigma_{\rm background }$, as a function of aperture diameter $D$ by measuring the flux scatter in blank-sky apertures on the background of the MIDIS Deep. Specifically, we compute the standard deviation of the flux distribution within apertures of different diameters and fit the resulting trend with a power-law. The relation is given by:

\begin{equation}
    \label{eqn:PowerLaw}
    \sigma_{\rm background} = \sigma_{\rm background}^{\rm per~pixel}\alpha D^\beta
\end{equation}

where $\sigma_{\rm background}^{\rm per~pixel}$ is the standard deviation of the flux distribution per non-masked pixel, $D$ the aperture diameter in arcseconds, $\alpha$ is a normalization factor, and $\beta$ is the power-law index. The best fit of $\sigma_{\rm background}$ as a function of aperture diameter provides the uncertainty to associate with the flux of a faint source measured in a given aperture. As expected, the empirical noise increases with aperture size, since larger apertures include more background noise (Fig.~\ref{fig:NoiseStudyFig}, Appendix~\ref{sec:AppendixB_no_mask} for the data table).

\subsubsection{MIRI noise from the pipeline variance map}
\label{sec:MIRInoisePipeline}

We compute the variance-based errors to compare with the empirical errors. We measure the average error, $\sigma_{\rm \sqrt{VAR}}$, for a given aperture diameter calculated from the variance image. We obtain $\sigma_{\rm \sqrt{VAR}}$ by summing the pixels in the same non-masked apertures as for obtaining $\sigma_{\rm background}$ in the variance image associated with the science image and taking the average of all sums. We then fit these data points with the following linear function:

\begin{equation}
    \label{eqn:Linear}
    \sigma_{\rm \sqrt{VAR}} = \sigma_{\rm \sqrt{VAR}}^{\rm per~pixel} + a D 
\end{equation}

Here $\sigma_{\rm \sqrt{VAR}}^{\rm per~pixel}$ is the average error in non-masked pixels in the variance image, $a$ is the slope, and $D$ is the aperture diameter in arcseconds. The average errors increase linearly with aperture size.

\subsubsection{MIRI noise for MIRI-only sources }
\label{sec:MIRInoiseMIRIonly}

In order to compare the noise from the background from the empirical method (Sect.~\ref{sec:MIRInoiseBackground}) with that obtained from the variance image (Sect.~\ref{sec:MIRInoisePipeline}), we fit the linear function in Eq.~\ref{eqn:Linear} to the empirical noise data points ($\sigma_{\rm background}$). We then compute the ratios of the best-fit slopes, $\eta$, between the empirical and the variance fits: 
\begin{equation}
    \label{eqn:Slope}
    \eta=\frac{a}{a_{\rm background}}
\end{equation}

where $a$ is the best-fit slope to $\sigma_{\rm \sqrt{VAR}}$ as a function of $D$, and $a_{\rm background}$ is the corresponding slope for $\sigma_{\rm background}$. The ratio of the best-fit slopes, $\eta$, is the rescaling factor to apply to the variance image ($\sqrt{\eta}$ to the error image) when deriving photometric errors with e.g., \textsc{SourceExtractor}. We measure $\eta\,=\,3.1\pm0.2$, meaning the errors of faint detections in MIDIS Deep are underestimated by a factor of three if we use a standard photometric tool. 

As a consistency check, we rescale the variance image by the derived factor and recompute the average errors ($\sigma_{\rm \sqrt{VAR resc}}$), to confirm that they agree with the empirical errors ($\sigma_{\rm background}$). The linear best fits to $\sigma_{\rm background}$ and $\sigma_{\rm \sqrt{VARresc}}$ are consistent, suggesting that the rescaling factor ($\eta$, Table~\ref{table:NoiseResults}) provides a good approximation of the empirical errors measured from the background.

\begin{figure}[h] 
   \centering
  \includegraphics[width=\hsize]{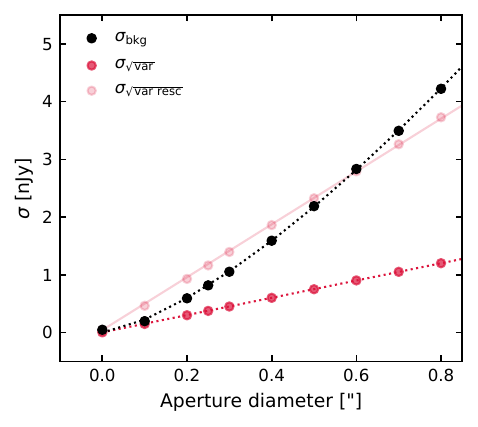}
   \caption{ Background noise level as a function of aperture diameter of MIDIS Deep. The black markers indicate the background noise level derived from the scatter in empty background apertures. The red markers denote the average error derived from the sum in quadrature of the pixels in the associated variance map to the MIDIS observations. The shaded red markers denote the average error derived after rescaling the variance maps. The background noise levels ($\sigma_{\rm bkg}$) are used as the 1$\sigma$ error for photometric measurements in given aperture sizes, as the average errors ($\sigma_{\rm var}$) are underestimated.  }

\label{fig:NoiseStudyFig}
\end{figure}

We summarize the best-fit values to Eqs.~\ref{eqn:PowerLaw},~\ref{eqn:Linear}, and~\ref{eqn:Slope} in Table~\ref{table:NoiseResults}. The errors with the rescaling method are slightly overestimated for small apertures (about $<0\farcs{4}$) and slightly underestimated for large apertures (about $>0\,\farcs{4}$) due to deviation from linearity as a consequence of correlated noise.  As expected, the smaller pixel scales also underestimate more the uncertainties. In Fig.~\ref{fig:NoiseStudyFig}, we plot the $\sigma_{\rm background}$, $\sigma_{\rm \sqrt{VAR}}$, and $\sigma_{\rm \sqrt{VARresc}}$ as a function of aperture size. The resulting values for the MIDIS Deep observations across different pixel scales and reduction versions are in Appendix \ref{sec:AppendixA_MIDIS_versions}.

For the remainder of this paper, we adopt the empirical errors as the uncertainties associated with our photometric flux measurements; the corresponding aperture corrected $1\sigma$ errors are listed in Table~\ref{table:DepthTable}.

\begin{table}[h]
\caption{ Best-fit parameters to Eqs.~\ref{eqn:PowerLaw},~\ref{eqn:Linear}, and~\ref{eqn:Slope}. When using the variance maps in MIDIS Deep, $\eta$ is the rescaling factor applied to the total variance map (and $\sqrt{\eta}$ to the error map) to obtain accurate photometric uncertainties. }
\label{table:NoiseResults}
\centering
\begin{tabular}{  c| l|ll|}
\hline \hline

 Equation & Parameters & Best-fit values \\\hline
(1) $\sigma_{\rm background} = \sigma_{\rm background}^{\rm per~pixel}\alpha D^\beta$ & $\sigma_{\rm per pixel}$ &0.049     \\
& $\alpha$ &   117.7 $\pm$ 0.4  \\
& $\beta$ &  1.41 $\pm$ 0.01     \\  \hline

& $\mu_{\rm per pixel}$ &  0.005   \\
(2) $\sigma_{\rm \sqrt{VAR}} = \sigma_{\rm \sqrt{VAR}}^{\rm per~pixel} + a D$  & $a$ &   1.50 $\pm$ 0.01 \\
& $a_{\rm background}$ &    4.6 $\pm$ 0.2  \\ \hline

(3) $\eta=\frac{a}{a_{\rm background}}$ & $\eta$ & 3.1 $\pm$ 0.2    \\

\hline 
\end{tabular}
\end{table}

\subsection{Catalog for MIRI-red sources: MIDIS Deep detections and NIRCam counterparts}
\label{sec:MIDISdeepCatalog}

We construct a catalog with aperture fluxes optimized to find MIRI-red sources among the 1682 MIRI detections in MIDIS Deep, excluding the bright sources (brighter than 23.5 mag at 5.6\,$\,\mu$m) and the detected sources in the masked region around them (Sect.~\ref{sec:StellarMask}). We extracted sources in MIDIS Deep with the python encoded \textsc{SourceExtractor}\footnote{thresh=1.5, var=img\_dict['var'], minarea=9, filter\_type='matched', deblend\_nthresh=32, deblend\_cont=0.0005, clean=True, clean\_param=1.0} package \citep{Bertin1996,Barbary2016_SEP_code} after smoothing the F560W image with a $3\times3$ Gaussian kernel with $\sigma\,=\,1.5$. We detected the NIRCam sources on an inverse-variance weighted image, combining F277W, F356W, and F444W, with the same detection parameters as for F560W, and detect 2082 sources in MIDIS Deep (also excluding the masked regions). In order to estimate the purity and the completeness of the MIRI detection (Sect.~\ref{sec:MIDISdeepPurityCompleteness}), we added to our MIRI catalog whether a MIRI detection has one or more NIRCam matches within a search radius of 0\farcs25, and applied the same procedure to the NIRCam catalog to identify sources with MIRI matches. 

We then measured, for both catalogs, the fluxes in MIDIS and in all the NIRCam broadbands in circular apertures of 0\farcs{4} of diameter after an additional background subtraction to account for possible residual background effects. We applied an aperture correction to get the total flux of the sources, compensating for the flux loss because of the PSF. We calculated the aperture correction factors from the \href{https://webbpsf.readthedocs.io/en/latest/}{\textsc{WebbPSF} python package} version 1.2.1 \citep{Perrin2012docu,Perrin2012,Perrin2014}. As mentioned previously, we adopt the empirical error measured in Sect.~\ref{sec:MIRInoise} for an aperture diameter of 0 \farcs 4 (Fig.~\ref{fig:NoiseStudyFig} and Table~\ref{table:DepthTable}).

We last compile and add to the catalogs the spectroscopically confirmed sources compiled in the region of MIDIS Deep (see Sect.~\ref{sec:OtherData}).

\subsection{MIDIS Deep purity and completeness}
\label{sec:MIDISdeepPurityCompleteness}

In this subsection, we assess the purity and completeness of the MIDIS Deep detections using the MIRI and the NIRCam catalogs. This analysis allows us to determine the purity and the completeness at $5\sigma$, the magnitude cut for selecting MIRI-red galaxy candidates (Sect.~\ref{sec:SelectionCriteria}), and to estimate the expected number of MIRI-red sources in MIDIS Deep based on several luminosity functions from the literature (Sect.~\ref{sec:NumberDensity}). We assume the NIRCam detections as true, given the NIRCam detection image is significantly deeper than the MIDIS Deep observations (see discussion in Sect.~\ref{sec:Depths} and Table~\ref{table:DepthTable}). 

For the purity, we binned the MIRI detections as a function of F560W S/N, from 1 to 30 with a bin width of 1. In each bin, we counted the number of MIRI detected sources with one or more NIRCam match. We obtained the purity per bin by dividing the number of MIRI detections with one or more NIRCam matches by the total number of MIRI detections and calculate binomial uncertainties for the fraction in each bin. We fitted the purity curve with the sigmoid-type function in \cite{ASPECS1_Walter2016}, used to fit the purity of ASPECS, whose parametrization is:

\begin{equation}
    \label{eqn:erf}
    P({\rm S/N}) = \frac{1}{2}\operatorname{erf}\left(\frac{ {\rm S/N} - C}{\sigma} \right) + \frac{1}{2}
\end{equation}

We obtain best fit values of $C\,=\,1.24\pm3.61$ and $\sigma\,=\,0.65\pm1.03$. We retrieve the F560W magnitudes for different purity levels from the inverse function of Eq.~\ref{eqn:erf}:
\begin{equation}
    \label{eqn:inv_erf}
    {\rm S/N}(P) = C + \sigma \operatorname{erf}^{-1}(2P-1)
\end{equation}

As for the purity, we utilize the MIRI and NIRCam detection catalogs to quantify the completeness of the detections as a function of MIRI magnitude and flux. The completeness in each bin is the fraction of NIRCam detections that match at least one MIRI detection ($\geq5\sigma$ at 5.6\,$\mu$m) out of all NIRCam detection in that bin. We bin the NIRCam detections with respect to their magnitude in F444W from mag 27 to mag 33 in 50 bins, bin width of 0.12, and compute binomial errors as well. We fitted the purity with a sigmoid-type function of magnitude, whose parametrization is:

\begin{equation}
    \label{eqn:erf_compl}
    C(m) = \frac{1}{2} \left(1 - \operatorname{erf}\left(\frac{m - m_{50}}{(\sqrt2 \sigma} \right) \right)
\end{equation}

where $c_{50}$ is the F560W magnitude at which the completeness reaches 50\%, and $\sigma$ is a width parameter that controls the steepness. We obtain best fit values of $m_{50}\,=\,28.83\pm0.01$ and $\sigma\,=\,0.85\pm0.01$. We retrieve the F560W completeness for different purity levels from the inverse function of Eq.~\ref{eqn:erf_compl}:

\begin{equation}
    \label{eqn:inv_erf_compl}
    m(C) = m_{50} + \sqrt2  \sigma \operatorname{erf}^{-1}(1 - 2C)
\end{equation}

In Table~\ref{tab:PurityCompleteness}, we report F560W magnitudes, purities, S/N, and completeness levels corresponding to purities ranging from 50 to 99\% and at the $5\sigma$ level in MIDIS Deep. Relaxing the purity threshold increases the statistical sample size, but at the cost of having detections with even lower F560W S/N. At $5\sigma$ in MIDIS Deep, we estimate a purity of detections of 92\% and completeness level of 54\%.

\begin{table}[H]
\caption{ Measured magnitudes, number of MIRI-red candidates, F560W S/Ns, and completeness levels and  for different levels of purities in MIDIS Deep. At a $5\sigma\,=\,28.75$ detection threshold, the survey reaches 92\% purity and 54\% completeness. }
\label{tab:PurityCompleteness}
\centering
\begin{tabular}{ c c c c c}
\hline \hline

P [\%] & 5.6\,$\mu$m mag & \# candidates & S/N  & C [\%]\\ \hline

50 & 30.22 & 35 & 1.28 & 5.21 \\
60 & 29.79 & 34 & 1.91 & 13.18 \\
70 & 29.46 & 27 & 2.59 & 23.12 \\
75 & 29.32 & 22 & 2.96 & 28.64 \\
80 & 29.17 & 14 & 3.37 & 34.56 \\
90 & 28.87 & 8 & 4.47 & 48.38 \\
\textbf{92} & \textbf{28.75} & \textbf{7} & \textbf{5.00} & \textbf{54.09} \\
95 & 28.67 & 5 & 5.37 & 57.65 \\
99 & 28.37 & 2 & 7.06 & 70.58 \\

\hline
\end{tabular}
\end{table}

\begin{figure}[H] 
   \centering
  \includegraphics[width=\hsize]{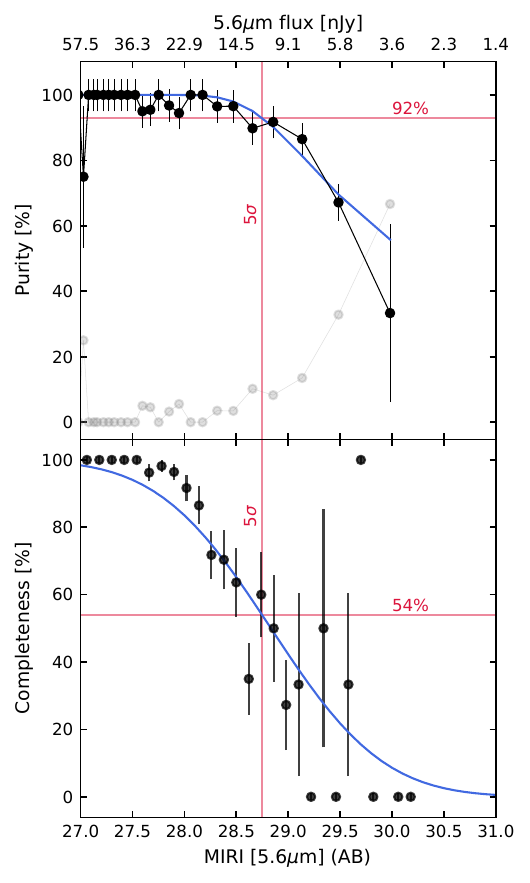}
   \caption{ Purity and completeness of the MIDIS Deep survey (excluding detection brighter than 23.5 mag) as a function of MIRI 5.6\,$\mu$m magnitude and flux. At a $5\sigma$ detection threshold, the survey reaches 92\% purity and 54\% completeness.
   }
\label{fig:MIRIpurity}
\end{figure}

Fig.~\ref{fig:MIRIpurity} shows the purity and best fit as a function of F560W MIRI magnitude in the top panel and the completeness in the bottom panel. The faint end (>30 mag) of the purity space is naturally limited by the detection limit.

\subsection{MIRI-red selection criteria}
\label{sec:SelectionCriteria}

We define selection criteria for MIRI-red galaxy candidates in our MIRI catalog (Sect.~\ref{sec:MIDISdeepCatalog}) making use of the empirical noise and depth measurements (Sect.~\ref{sec:MIRInoise}) and the limiting magnitude of $5\,\sigma$ reaching 92\% purity and 54\% completeness in the MIRI detections (Sect.~\ref{sec:MIDISdeepPurityCompleteness}). We select MIRI-red galaxy candidates as $5\sigma$-detected in F560W and with colors of $m_{\rm F444W}\,-\ m_{\rm F560W} \geq 0.5$. The MIRI-red galaxy candidate selection criteria are:

\begin{equation}
\label{eqn:MIRIselection}
\begin{split}
m_{\rm F560W} &\leq m_{\rm 5\sigma,F560W} = 28.75 \\
m_{\rm F444W} &- m_{\rm F560W} \geq 0.5 \\
\end{split}
\end{equation} 

MIRI-red sources are detected down to 29.25 mag in F444W. Sources fainter than 29.25 mag are undetected in F444W, corresponding to possible MIRI-only sources and implying $m_{\rm F444W}\,-\,m_{\rm F560W} >2.0$. Undetected at $<1\sigma$ in NIRCam broadbands is a strict threshold, but since we estimate the depth in larger apertures compared to the usual adopted apertures size it is reasonable. Our 1$\sigma$ limit corresponds in reality to about 3-5$\sigma$ for depth measured in apertures optimized for the shorted wavelength observations (see discussion in Sect.~\ref{sec:Depths}).

\section{Results}
\label{sec:Results}

In the following subsections, we present the candidates for MIRI-red galaxies in MIDIS Deep (Sect.~\ref{sec:ResultsMIRICandidates}) and characterize galaxy populations with red F444W$-$F560W colors in MIDIS Deep (Sect.~\ref{sec:ResultsMIRIonlyCharachterization}).

\subsection{MIRI-red galaxy candidates}
\label{sec:ResultsMIRICandidates}

\begin{figure*}[h!] 
\centering
   \includegraphics[width=1.0\hsize]{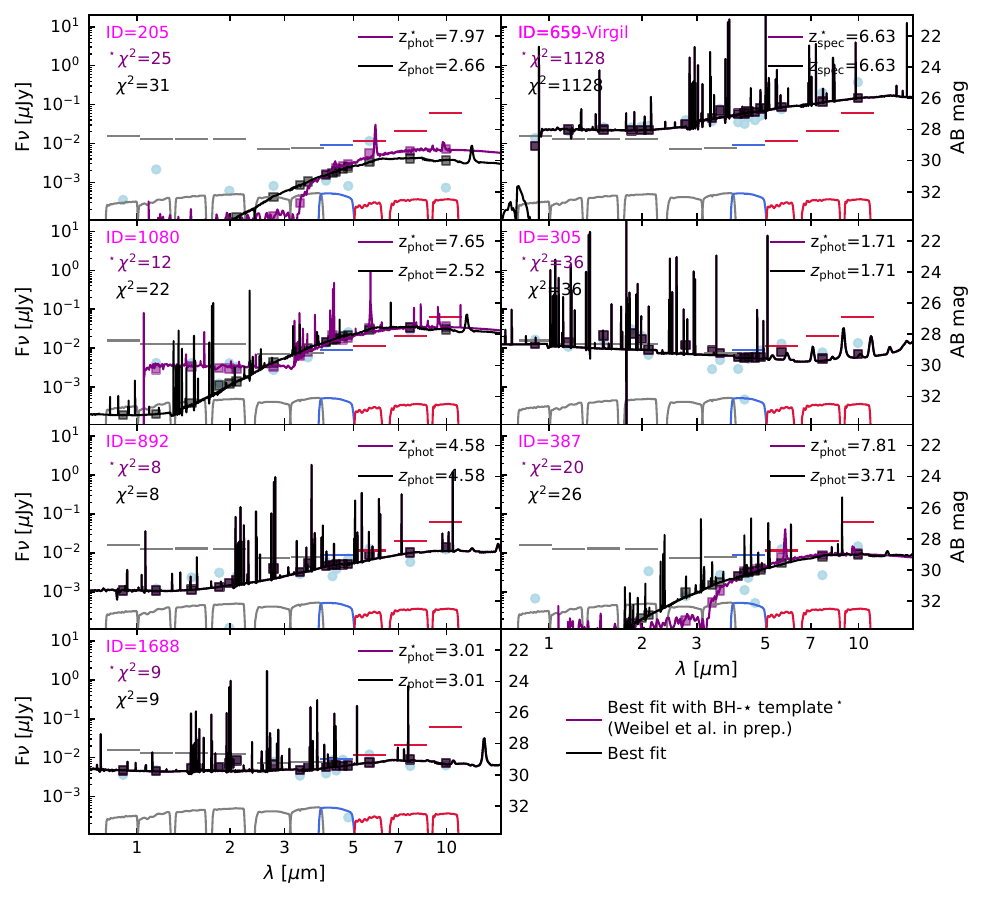}
      \caption{ Eazy SED fit of the MIRI-red galaxy candidates sample including the deep F770W and F1000W data from MIDIS-red. In Appendix \ref{sec:AppendixD_SED}, we show the same figure without MIDIS-red data. Interestingly, without the MIDIS-red data, MIRI-red candidate with ID=205 shows H$\alpha$ emission line in F560W instead of F770W, as it is not constrained without the MIDIS-red observations. In Appendix \ref{sec:AppendixD_SED} are the photometric redshift probability distribution corresponding to all the fits. }
         \label{fig:MIRIonly_red_SED}
\end{figure*}

Following the selection criteria in Eq.~\ref{eqn:MIRIselection}, we detect seven MIRI-red candidates. Among these, one source has S/N$<$1 in all NIRCam broad bands and is therefore classified as a MIRI-only candidate. In Appendix~\ref{sec:AppendixC1_Cutouts}, we provide cutouts of the seven sources in all NIRCam broadbands, F560W, F770W, and F1000W, and indicate the S/N in every band (see Fig.~\ref{fig:LETTER_MIRIonlyFigure1} for the location of the seven sources in MIDIS Deep). We list the catalog IDs, coordinates, F444W, F560W, F770W, and F1000W fluxes, and F444W-F560W colors of the MIRI-red galaxy candidates in Table~\ref{tab:BothMIRIonlyTab}.
Figure~\ref{fig:MIRIonly_red_SED} shows the SED fits of the sample with \textsc{EAZY} \citep{Brammer2010_eazy_soft} including the deep F770W and F1000W MIDIS-red observations (see Sect.~\ref{sec:MIDIS}). In Appendix~\ref{sec:AppendixD_SED} are the best fits only including the $\sim10$\,h in F1000W from MIDIS and the photometric redshift probability distributions of all the fits. We perform the fits twice using the approach of Weibel et al. in prep., meaning fitting the 13 \textsc{EAZY} templates generated from Flexible Stellar Population Synthesis (FSPS) models \citep{Larson2023_eazytemplates} complemented with the blue template (created to model a strong emission line galaxy at $z=8.5$ \citep{Carnall2023_template}) with and without including a BH-$\star$ template. Weibel et al. in prep. create a BH-$\star$ template from the NIRSpec/PRISM spectrum of \textit{The Cliff} and extend it to longer wavelength with a blackbody fit. In Fig.~\ref{fig:MIRIonly_red_SED}, we show the fits with the BH-$\star$ template in purple and denote it with the symbol $^\star$. These fits show a preferred solution with the BH-$\star$ template. The solutions without the BH-$\star$ template suggest a dusty galaxy with a very low stellar mass, $\log M_\star/M_\odot\sim7$. However, \citet{Popping2020} perform a deep search for dust continuum in ASPECS, the ALMA survey overlapping with the MIDIS Deep field, and report dusty galaxies only up to $z=2$ and with a dust mass of $M_{\rm dust}=10^8$ M$_\odot$. A low-mass and dusty solution for MIRI-red sources is thus ruled out.

We report the MIRI-red galaxy candidate (catalog ID=205) with S/N$_{\rm\,F560W}$=5.0, and less than 1$\sigma$ in all NIRCam broadbands, as a candidate for a MIRI-only source. This source, however, is at the intersection of the PSF spike extend from two bright sources, beyond the spatial extent of the PSF models used in the masking procedure (Sect.~\ref{sec:StellarMask}). Therefore, this source is considered as likely spurious. Among the six other MIRI-red sources, there is the extended LRD in MIDIS \textit{Virgil} (catalog ID=659) \citep{Iani2025_Virgil,Rinaldi2025_virgil} with a MUSE spectroscopic redshift of $z=6.63$. Notably, we have a F115W dropout (catalog  ID=1080, F090W-F115W$\geq0.3$ mag), a high-$z$ galaxy candidate with no spectroscopic observation to date. The source with ID=387 is spurious, it is beyond the coverage of the PSF mask (Sect.~\ref{sec:StellarMask}).

\begin{table*}[h!]
\caption{ List of the MIRI-red galaxy candidates (Eq.~\ref{eqn:MIRIselection}). All the fluxes are in units of nJy. $^a$ MUSE spectroscopic redshift \citep{Iani2025_Virgil}.  }
\label{tab:BothMIRIonlyTab}
\centering
\begin{tabular}{c|c|c|c|c|c|c|c|c}
\hline \hline
 ID  & ra [deg] & dec [deg] & $F_{\rm F444W}$ & $F_{\rm F560W}$ & $F_{\rm F770W}$ & $F_{\rm F1000W}$  & F444W$-$F560W & $z_{\rm spec}$\\ 
\hline

 205 & 53.157563 & -27.796503 & -0.39 & 11.59 & -0.89 & 0.73(20.18) & >2.0 & -- \\ 
659 & 53.158056 & -27.786313 & 47.17 & 76.17 & 201.19 & 377.81(380.30) & 0.52 & 6.63$^a$ \\
 1080 & 53.172795 & -27.783147 & 18.66 & 33.17 & 27.81 & 34.50(25.95) & 0.62 & --   \\ 
305 & 53.157726 & -27.794213 & 5.32 & 13.07 & 7.81 & 13.55(15.31) & 0.98  & --  \\  
892 & 53.154892 & -27.779249 & 3.78 & 12.84 & 5.85 & 12.30(-2.66) & 1.33  & --  \\  
387 & 53.160937 & -27.793471 & 2.30 & 12.53 & 2.69 & 14.76(22.84) & 1.84 & --  \\  
1688 & 53.185038 & -27.777380 & 4.66 & 12.06 & 6.10 & 6.31(-8.67) & 1.03 & --   \\ 
\hline

\end{tabular}
\end{table*}

\begin{figure*}[t] 
   \includegraphics[width=\hsize]{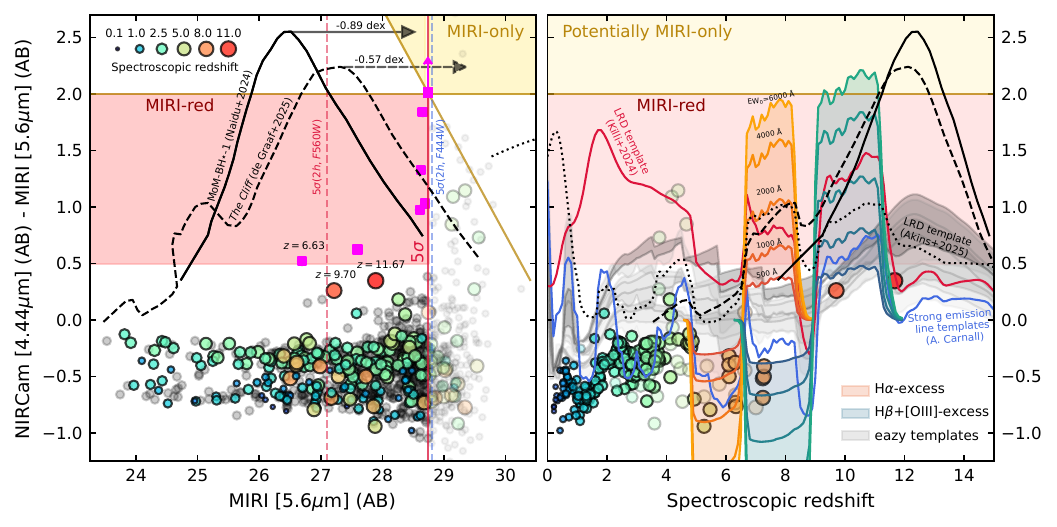}
      \caption{NIRCam/F444W$-$MIRI/F560W colors of MIRI detections in the MIDIS Deep survey with F560W magnitudes $>23.5$\,mag, shown as a function of F560W magnitude in the left panel and spectroscopic redshift in the right panel. The red and golden shaded region indicates the color space occupied by MIRI-red and potential MIRI-only sources.} 
         \label{fig:ColorPlot}
\end{figure*}

\subsection{Characterization of MIRI-red galaxies in MIDIS}
\label{sec:ResultsMIRIonlyCharachterization}

\begin{figure*}[t] 
   \includegraphics[width=\hsize]{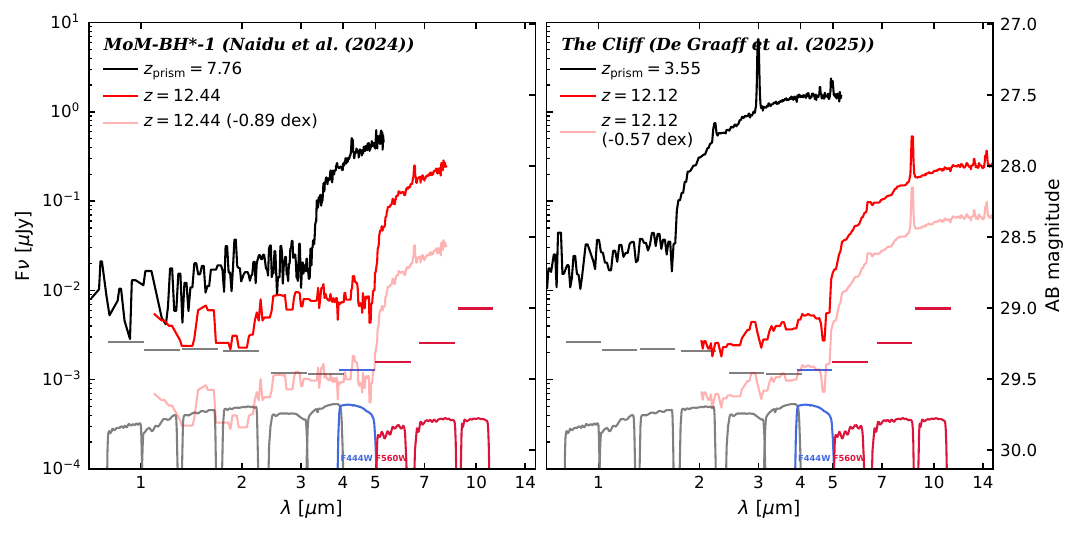}
      \caption{ Spectra of MoM-BH$\star$-1 \citep{Naidu2025_mom} and \textit{The Cliff} \citep{DeGraaff2025_thecliff}, shown at their observed redshifts (in black), and at the redshift where they appear reddest in F444W$-$F560W (in red, see also Fig.~\ref{fig:ColorPlot}), and scaled to a $5\sigma$ detection in F560W (in light red).} 
         \label{fig:SEDs}
\end{figure*}

To understand the type of galaxies that fulfill the MIRI-red criteria (Eq.~\ref{eqn:MIRIselection}), we look at all the MIRI Deep detection in a plot of F444W-F560W color as a function of F560W magnitude. Fig.~\ref{fig:ColorPlot} shows the F444W$-$F560W colors of the 1040 MIRI-detections in MIDIS Deep. This excludes any source brighter than 23.5 mag as we use the mask explained in Sect.~\ref{sec:StellarMask}. Thus, none of the MIDIS-ALMA sources from \citet{Boogaard2023} are shown in this color plot, as they are all in the range 19.97 to 23.42 mag in F560W. We color the 206 (>23.5 mag in F560W) galaxies by their spectroscopic redshifts, and show any other detection in black. The two noticeable tracks in colors are a consequence of the stellar bump at 1.6\,$\mu$m rest-frame \citet{Sawicki2002_stellar_bump}, which falls into the F560W filter at $z=2.2$, when observed at 5\,$\mu$m, between F444W and F560W. The scatter in the F444W$-$F560W colors increases at fainter detections, especially below the 5$\sigma$ limit, where the flux measurements become noisy. We derive an average F444W-F560W color for the higher-redshift branch of $-$0.3, that we used in the determination of the MIRI detection completeness in Sect.~\ref{sec:MIDISdeepPurityCompleteness}. The red and gold regions in Fig.~\ref{fig:ColorPlot} indicate the parameter spaces of MIRI-red galaxy candidates and potential MIRI-only sources in MIDIS.

None of the 206 spectroscopic galaxies in MIDIS Deep exceeds a F444W$\,-\,$F560W color of $0.5$ mag, as we define for our MIRI-red sample (Eq.~\ref{eqn:MIRIselection}). Among the spectroscopic galaxies with the reddest F444W$-$F560W colors, we find the two high-$z$ galaxies JADES-GS-z9.7 ($z=9.7$) and JADES-GS-z11-0 ($z=11.67$) \citep{CurtisLake2023_z11REF,Bunker2023_GNz11_jwst,Hainline2024_z11REF}, that will be extensively discussed in Melinder et al. in prep. with F444W-F560W>0.25 mag. Both of these galaxies have strong H$\beta$+[OIII] emission that boosts the flux in F560W. The spectroscopic red F444W-F560W galaxies in MIDIS seem to have an excess of flux in F560W from a strong emission line. The proximity of F444W and F560W and the survey design and depth (>28 mag) naturally selects these red galaxies as high redshifts rather than intermediate redshifts, massive and dusty or AGNs as shallower observations with MIRI \citep{Wang2023,Alberts2024_SMILES,Lyu2024_MIRI,Kakkad2025_MIRI,RamosAlmeida2025_MIRI}.

To identify sources that could fulfill the MIRI-red criteria (Eq.~\ref{eqn:MIRIselection}) and even reach the parameter space of MIRI-only sources, we enlarge the search of observed galaxies beyond MIDIS Deep with NIRSpec spectroscopy in the \textit{DJA}, we identify two LRDs with extreme F444W$-$F560W colors if observed at higher-$z$: \textit{The Cliff} and MoM-BH$\star$-1. \textit{The Cliff} (selected in the NIRCam PRIMER Survey, GO-1837, PI: Dunlop, and RUBIES-UDS-154183, \citet{DeGraaff2025_thecliff}) is a galaxy at $z_{\rm spec}=3.548$ with a Balmer break of BB$=6.9^{+2.8}_{-1.5}$. MoM-BH$\star$-1 (\citet{Naidu2025_mom}, JWST Cycle 3 program GO-5224 PIs: Oesch \& Naidu) is a galaxy at $z_{\rm spec}=7.7569$ with an exceptional Balmer break of BB=$7.7^{+2.3}_{-1.4}$. Their predicted F444W$-$F560W color (Fig.~\ref{fig:ColorPlot}, dashed and solid black lines) enter the MIRI-only color space at $z\sim11\,-\,13.5$, reaching colors as extreme as F444W$-$F560W=2.24 (\textit{The Cliff} at $z=12.12$, F444W = 29.56 mag, F560W = 27.32 mag) and F444W$-$F560W=2.55 (MoM-BH$\star$-1, $z=12.44$, with F444W = 29.07 mag, F560W = 26.52 mag). Both would satisfy the MIRI-red criteria and be undetected in F444W if dimmed by $\sim$0.75 mag, e.g., due to lower stellar masses. Fig.~\ref{fig:SEDs} shows their spectra scaled to an S/N of 5 in MIDIS Deep (requiring an extra -0.57 dex and -0.89 dex, respectively), the observed spectra and the unscaled spectra. In Fig~\ref{fig:ColorPlot}, we show the color tracks with redshift of \textit{The Cliff} and MoM-BH$\star$-1.

Next we explore which galaxy templates populate the MIRI-red and MIRI-only color space in Fig.~\ref{fig:ColorPlot} and are the reddest at 5.6\,$\mu$m. We adopt the templates set \textsc{blue\_sfhz} from \textsc{EAZY} \citep{Brammer2010_eazy_soft,Larson2023_eazytemplates}, which includes 13 FSPS \textsc{EAZY} templates, complemented with two strong emission line galaxy templates: the red template (to fit an $z\,=\,4.53$ LRD \citep{Killi2024_redtemplate}) and the blue template (to model a strong emission line galaxy at $z=8.5$ \citep{Carnall2023_template}). We also include the averaged LRD template from \citet{Akins2024_LRD_template}, built from the best-fit SEDs of 434 photometric LRDs combining AGN and galaxy components, see Fig.~\ref{fig:ColorPlot}. MIRI-red galaxies can be: a $z\sim2$ LRD, a massive and dusty $z\sim3$ galaxy, a strong emission line galaxy, or a high redshift evolved galaxy. However, none of the template reaches F444W$-$F560W$\,>\,2.0$ (the MIRI-only color space). Only MoM-BH$\star$-1 and \textit{The Cliff} would populate the 5.6\,$\mu$m MIRI-only color space if these two galaxies were fainter and at higher redshifts. These results suggest that future galaxy models are needed to explain potential MIRI-only galaxies and galaxies such as MoM-BH$\star$-1 or \textit{The Cliff}. MIRI-only galaxies at 5.6\,$\mu$m are very extreme, and the proximity between F444W and F560W constrains the galaxy to show an extreme continuum break or an emission line, as opposed to dust reddening or dust emission for example, that are expected in large area and shallower surveys. Consistently with our SED fits in Fig.~\ref{fig:MIRIonly_red_SED}, the solutions with the \textsc{EAZY} templates with and without the BH-$\star$ template are \textit{The Cliff} and a moderate redshift and dusty galaxy.

\section{Discussion}
\label{sec:Discussion}

\subsection{On the nature of MIRI-red sources: possible physical origins}
\label{sec:OnTheNature}

We discuss the possible nature of MIRI-red galaxy candidates within the limits defined by the selection criteria and the depth of MIDIS Deep. We further extend the discussion to the F444W$-$F560W color in the case of a MIRI-red source that is undetected in F444W and may be a MIRI-only source if it is also undetected at shorter wavelengths. 

Placing the rest-frame UV continuum break at $\lambda_{\rm obs}\,=\,5.1$\,$\mu$m, the wavelength enclosing 10\% of the total flux in the MIRI/F560W bandpass, we derive a redshift of $z_{\rm Lyman ~scenario}=41.41$, only 61.5 Myr after the Big Bang for a 5.6\,$\mu$m dropout. From the $5\sigma$ limiting magnitude in F560W, as an upper limit on the apparent magnitude, the distance modulus, and applying the $K$-correction at $z_{\rm lyman ~scenario}$, we measure a lower limit on the absolute UV magnitude of $M_{UV}>-20.7$.

Compared to the $M_{UV}$ of the known highest-$z$ galaxies, the limit we derive is at least not brighter. The three furthest observed galaxies to-date are MoM-z14 \citep{Naidu2025_MoM_z14}, JADES-GS-z14-0, and JADES-GS-z14-1 \citep{Carniani2024_JADES-GS-z14-0-1} and have $M_{UV}=-20.2$, $M_{UV}=-20.81$, and $M_{UV}=-19.0$, respectively. Although pushing the dropout technique into the wavelength range covered by MIRI could be the next breakthrough in the field of high-$z$ galaxies, we do not expect galaxies at such high-$z$, within the theoretical framework of the $\Lambda$CDM cosmological model \citep{Planck2016,Planck2020}, where the first galaxies formed around $z\sim20$ and the first stars around $z\sim30$ \citep{Bromm2004,Bromm2011_first_gal_Review}. 

Similarly, by locating the Balmer break at $\lambda_{\rm obs}=5.1~\mu$m, we obtain a galaxy redshift of $z_{\rm Balmer~scenario}=13.15$ and a Balmer break strength of $BB_{\rm MIRI-red}=1.6$, with NIRCam/F444W probing the flux bluewards and MIRI/F560W measuring the flux redwards of the break. Assuming a MIRI-only scenario (i.e., a source undetected in F444W), the Balmer break strength would be $BB_{\rm MIRI-only}=6.3$. Note that these are an upper limits, as the calculation assumes a continuum only with no emission line contamination. The Balmer break strength required for a source to be classified as MIRI-red is high but consistent with models and observations. In contrast, the Balmer break strength required for a MIRI-only source exceeds the range predicted by standard galaxy models, assuming a Balmer-break-only scenario at $z=13.15$ \citep{Steinhardt2023_Balmerbreak,Wilkins2024,Cullen2024,Narayanan2025}. Nevertheless, as discussed in Sect.~\ref{sec:ResultsMIRIonlyCharachterization}, unexpectedly strong Balmer breaks from LRDs have been observed at $z>7$ with MoM-BH$\star$-1 for example (see Fig.~\ref{fig:SEDs}).

For the scenario of H$\alpha$ or H$\beta$+[OIII] lines to boost the F560W flux, the galaxy lies at $z_{\rm H\alpha}\,=\,7.6$, $z_{\rm[OIII]5008}\,=\,10.3$, and $z_{\rm H\beta}\,=\,10.6$, (calculating the redshift with 5.6\,$\mu$m). We define a redshift range, $z_{\rm min}$ and $z_{\rm max}$, for which the line falls into F560W with the wavelength corresponding to the bandpass enclosing 10\% and 90\% of the total flux (see Table~\ref{tab:RequMIRIonly}). We estimate the rest-frame equivalent width (EW$_0$) for the emission lines to reach F444W-F560W$\geq$0.5 and F444W-F560W>2 magnitudes, the MIRI-red and the MIRI-only color spaces (Fig.~\ref{fig:ColorPlot}), respectively. From the emission line templates shown in Fig.~\ref{fig:ColorPlot}, see Appendix~\ref{sec:AppendixC_ExpNumber}, we obtain colors of F444W-F560W$\geq$0.5 and F444W-F560W>2 for templates with rest-frame equivalent widths of EW$_0({\rm H\alpha})\,=\,725$ Å, EW$_0{(\rm H\beta+[OIII]})\,=\,555$ Å, and EW$_0{(\rm [OIII]5008)}\,=\,370$ Å, and of EW$_0({\rm H\alpha})\,=\,6700$ Å, EW$_0{(\rm H\beta+[OIII]})\,=\,5100$ Å, and EW$_0{(\rm [OIII]5008)}\,=\,3415$ Å\footnote{We assume the average [OIII]+H$\beta$ observed line ratios reported in \cite{Meyer2024_OIIILF_fresco}.}. 

High-resolution hydrodynamic simulations predict that the extreme EWs to observe F444W$-$F560W>2 are not achievable. For example, FLARES reports EW$_0({\rm H\alpha})\lesssim700$ Å and EW$_0({\rm H\beta+[OIII]})\lesssim3100$ Å under the most extreme star-forming conditions, without invoking AGN activity \citep{Vijayan2021_FLARES,Wilkins2023_FLARES}. Photometric studies report larger values, with EW$_0({\rm H\alpha})\sim2300$ Å \citep{Bollo2023_HaLF,Boyett2024_HbOIIIEW} and EW$_0({\rm H\beta+[OIII]})\sim3000$ Å \citep{DeBarros2019,Endsley2021_EWOIII,Matthee2023_HbOIIIEW,Bouwens2023_HUDFLF}. Empirical extrapolations from pre-\textit{JWST} measurements suggest EW$_0({\rm H\alpha})\sim400$ Å at $z\sim8$ \citep{Fumagalli2012_3DHST_Ha}, and early \textit{JWST} results predict EW$_0({\rm H\beta+[OIII]})\sim1300$ Å by $z\sim10.5$ \citep{Laporte2023}. In contrast, spectroscopic measurements consistently yield lower EWs. The MIRI-red galaxies, selected with F444W$-$F560W>0.5, exhibit equivalent widths comparable to the highest values measured spectroscopically. NIRSpec and grism studies typically find EW$_0$ of a few hundred angstroms \citep{Fumagalli2012_3DHST_Ha,Meyer2024_OIIILF_fresco}, with stacked NIRSpec spectra reporting EW$_0({\rm H\alpha})\sim900$ Å and EW$0({\rm H\beta+[OIII]})\sim1300$ Å \citep{Roberts-Borsani2024_lines}. With MIRI/MRS spectroscopy \cite{AlvarezMarquez2024_MACS1149_MRS,AlvarezMarquez2025_MIRIHa,PrietoJimenez2025_MIRIHa} measure EW$_0({\rm H\alpha})\sim800$ Å in early and bright galaxies. The most extreme spectroscopic measurement to date is EW$0({\rm H\beta+[OIII]})=4116\pm690$ Å in a lensed galaxy at $z_{\rm spec}=7.0435$ with $\log({\rm M}/M\odot)=7.70\pm0.24$. Overall, spectroscopically confirmed EWs remain well below those required to explain MIRI-only colors, and are instead consistent with the values observed in MIRI-red galaxies and shown in Fig.~\ref{fig:ColorPlot}. Nevertheless, strong ionizing or primevial sources such as Pop III stars or DCBHs explored with photoionization models predict that in theory such a source is able to produce the extreme EW$_0({\rm H\alpha}$. For example, if a Pop III star or a DCBH is surrounded by metal-enriched gas, then the extreme EW$0({\rm H\beta+[OIII]})$ are theoretically reproducible \citep{Nakajima2022_popIII_DBHC,Nakajima2025}. However, Pop III stars and DCBHs have not yet been observationally confirmed even with the \textit{JWST} \citep{Schauer2020,Venditti2023,Fujimoto2025,Fujimoto2025_arxiv}.

In Table~\ref{tab:RequMIRIred}, we summarize the numbers and limits derived for MIRI-red galaxy candidates as high-redshift Balmer break galaxies and H$\alpha$ or [OIII]$\lambda$5008 emission lines excess in F560W. In Table~\ref{tab:RequMIRIonly}, we summarize the numbers and limits derived for MIRI-only sources as Lyman break galaxies, Balmer break galaxies, and H$\alpha$ or [OIII]$\lambda$5008 emission lines excess in F560W.

\begin{table}[h]
\caption{Redshift ranges and spectral properties of MIRI-red galaxy candidates in MIDIS Deep with F444W$\,-\,$F560W>0.5. }
\label{tab:RequMIRIred}
\centering
\begin{tabular}{ c c c c c }
\hline \hline
\multicolumn{2}{c}{MIRI-red galaxies }   & $\bar{z}$ & $z_{\rm min}$ & $z_{\rm max}$    \\ \hline 
 Balmer break & 1.6 &  14.50  & 13.15 & 15.65 \\
 $ \rm EW_0{(\rm H\alpha)}$  &  725 Å &  7.61  &  6.86 & 8.25 \\
 $ \rm EW_0{(\rm H\beta+[OIII])}$ & 555  Å  &  10.46 & 9.61 & 11.13  \\
 $ \rm EW_0{(\rm H\beta})$ & 185 Å &  10.63 & 9.61 & 11.49 \\
 $ \rm EW_0{(\rm [OIII]5008)}$ & 370 Å  &  10.28 & 9.30 & 11.13 \\
\hline
\end{tabular}
\end{table}

\begin{table}[h]
\caption{Redshift ranges and spectral properties of sources in MIDIS Deep with F444W$\,-\,$F560W>2 measured from the limits in F560W as the 5$\sigma$ 28.75 (11.5 nJy) and the $1\sigma$ depth of 30.75 (1.81 nJy) in F444W. }
\label{tab:RequMIRIonly}
\centering
\begin{tabular}{ c c c c c }
\hline \hline
\multicolumn{2}{c}{MIRI-only sources }   & $\bar{z}$ & $z_{\rm min}$ & $z_{\rm max}$    \\ \hline 
 Lyman break & $M_{\rm UV}$ > -20.7 &  45.47 & 41.41 & 48.94  \\
 Balmer break & 6.3 &  14.50  & 13.15 & 15.65 \\
 $ \rm EW_0{(\rm H\alpha)}$  &  5645 Å &  7.61  &  6.86 & 8.25 \\
 $ \rm EW_0{(\rm H\beta+[OIII])}$ & 4306  Å  &  10.46 & 9.61 & 11.13  \\
 $ \rm EW_0{(\rm H\beta})$ & 508 Å  &  10.63 & 9.61 & 11.49 \\
 $ \rm EW_0{(\rm [OIII]5008)}$ & 2884 Å &  10.28 & 9.30 & 11.13 \\
\hline
\end{tabular}
\end{table}

\subsection{MIDIS Deep number counts and comparison to luminosity function predictions}
\label{sec:NumberDensity}

\begin{figure}[h] 
   \centering
  \includegraphics[width=\hsize]{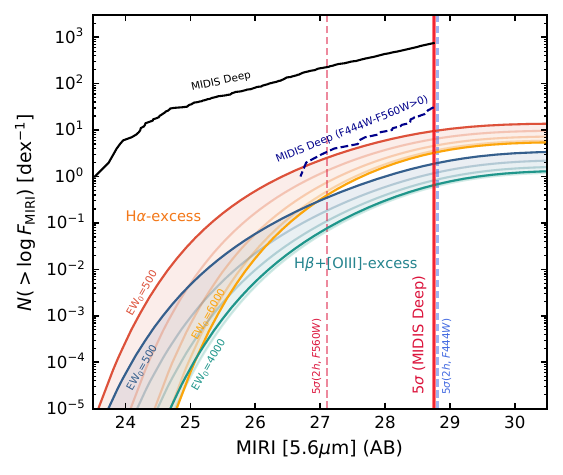}
   \caption{
   MIDIS Deep survey number counts for detections with F560W magnitudes $>23.5$ mag and aperture-corrected fluxes measured in apertures of diameter D\,=\,0\farcs{4}. These are compared to predicted number counts for H$\alpha$ and H$\beta$+[OIII] emitters derived from observed line luminosity functions from FRESCO and COSMOS-3D \citep{Covelo-Paz2025_HaLF_fresco,Meyer2025_OIIICOSMOS} for different values of emission-line rest-frame equivalent width.  }

\label{fig:NumberCounts}
\end{figure}

Figure \ref{fig:NumberCounts} shows the cumulative number of detections in MIDIS Deep as a function of magnitude (black line), along with the expected number of H$\alpha$ and H$\beta$+[OIII] line emitters in MIDIS Deep (shaded orange and teal regions). We utilize the observed spectroscopic line luminosity function measured in FRESCO \citep{Covelo-Paz2025_HaLF_fresco,Meyer2024_OIIILF_fresco} in F444W/Grism mode, just one band bluer than the MIRI/F560W filter. The methodology to calculate a mapping between line luminosity and observed flux in MIDIS and to obtain the cumulative number counts in MIDIS is explained in Appendix~\ref{sec:AppendixC_ExpNumber}. In Tables~\ref{tab:nexp_Ha} and~\ref{tab:nexp_HbOIII}, we summarize the expected number of H$\alpha$ and H$\beta$+[OIII] emitters in MIDIS Deep for EW$_0=\,500,\,1000,\,2000,\,4000,\,6000$ Å. The templates with EW$_0=\,500$ Å are slightly redder than the reddest spectroscopically confirmed sources in MIDIS in Fig.~\ref{fig:ColorPlot}. We expect 13 and 14 galaxies for H$\alpha$ and H$\beta$+[OIII] excess in F560W, respectively, with EW$_0=\,500$ Å. The number of H$\alpha$ excess galaxies is consistent with the finding from \cite{Rinaldi2023} who report 12 H$\alpha$ emitters with EW$_0$ ranging from 200 to 3000 Å, estimated from photometry only. Greve et al. in prep. report six galaxies with H$\beta$+[OIII] excess in F560W, finding a lower number of H$\beta$+[OIII] galaxies than expected from the extrapolated FRESCO luminosity function in F560W. If all MIRI-red galaxies were all H$\alpha$ or all H$\beta$+[OIII] emitters, it is half than the expected number. If one MIRI-red galaxy candidate was confirmed with EW$_0({\rm H\alpha})\sim6000$ Å or EW$_0({\rm H\beta+OIII})\sim4000$ Å , there would still be fewer detections than expected, as for those EW$_0$ we expect five galaxies for H$\alpha$ and four for H$\beta$+OIII.
Using the theoretical number density models for LRDs presented in \citet{Inayoshi2025}, we estimate the expected number of LRDs in the MIDIS survey volume (see Appendix~\ref{sec:AppendixC_ExpNumber} for details). Adopting the best-fit parameters and cosmology from \citet{Inayoshi2025} and from \citet{Tanaka2025_LRDphi}, we predict expected numbers of 0.05 and 0.01 LRDs, respectively. The low number of MIRI-red candidates in MIDIS is therefore fully consistent with these expectations. This highlights the intrinsic rarity of extreme LRDs such as MoM-BH$\star$ and \textit{The Cliff}, with only two such objects currently reported, only one of which lies at $z>6$.

We detect fewer emission lines galaxies (with EW$_0>500$ Å and F444W-F560W>0.0) than expected from extrapolating the FRESCO luminosity functions into F560W, suggesting that even if all the photometric MIRI-red candidates were spectroscopically confirmed, we would observe a decline of at least one half in the number density of H$\alpha$ and H$\beta$+[OIII] galaxies beyond $z>7$ and $z>9$, respectively. This suggests a turn in the physical process of galaxies at very high redshifts. Possible explanations to finding fewer than expected are the following. First, we assume no evolution with redshift in the FRESCO line luminosity functions extrapolated into the bluest MIRI band. A drop in density beyond $z>7$ for H$\alpha$ and for H$\beta$+[OIII] at $z>9$ is naturally expected from the drop of the CSFRD and the metallicity evolution with redshift in these redshift ranges. Second, the MIDIS Deep field is only 2.4 arcmin$^2$ and the HUDF is known for being an underdense region \citep{Uzgil2021_ALMA_CII,Helton2024}, thus the FRESCO line luminosity functions might not be representative of the emission line galaxies in MIDIS Deep. In our calculation for both estimates, we use the average luminosity functions from both GOODS-S, GOODS-N, and in the case of H$\alpha$ it includes data from the CONGRESS survey measured in F356W, at lower redshift. In the case of H$\beta$+[OIII], it is known that the GOODS-N field contains an overdensity at $z\sim7.1$ \citep{Fujimoto2022_overdensity} that might bias our first estimate of the expected number of [OIII] emitters in MIDIS Deep as MIRI-red galaxy candidates. We thus use the latest median [OIII] luminosity function from \cite{Meyer2025_OIIICOSMOS} in the 0.3 deg$^2$ covered by COSMOS-3D (a large NIRCam WFSS F444W progam, Program ID \href{https://www.stsci.edu/jwst/science-execution/program-information?id=5893}{5893}, PI K. Kakiichi, Kakiichi in prep.), 10 times the area of FRESCO, albeit two times shallower, that combined with other literature data constrains the knee of the [OIII] line luminosity function. \cite{Meyer2025_OIIICOSMOS} quantify a 2$\sigma$ overdensity in GOODS-N and reports that GOODS-S in perfect agreement with the median luminosity function therein, better constrained than the FRESCO one. With the COSMOS-3D LF, we estimate an expected number of three H$\beta$+[OIII] emitters in MIDIS Deep with EW$_0({\rm H\beta+[OIII]})=500$ Å and one with EW$_0({\rm H\beta+[OIII]})=4000$ Å (Table~\ref{tab:nexp_HbOIII_cosmos}). The latter result is consistent with the number of MIRI-red galaxy candidates. In Appendix \ref{sec:AppendixC_ExpNumber}, we show a version of Fig.~\ref{fig:NumberCounts} with the latest OIII luminosity function from \citep{Meyer2025_OIIICOSMOS}. Last, in the case of the H$\alpha$ line luminosity function, a fraction of the H$\alpha$ are AGNs. \cite{Covelo-Paz2025_HaLF_fresco} removes the AGN contamination and \cite{Matthee2024}, combining EIGER (Program ID \href{https://www.stsci.edu/jwst/science-execution/program-information?id=1243}{1243}, PI S. Lilly; \cite{Kashino2023_EIGERsurvey}) and FRESCO, estimate the Ha broad line AGN number density and report a significant fraction of H$\alpha$ emitters are broad line AGNs. One example in MIDIS is the broad line AGN \textit{Virgil} that is among the reddest detection in MIDIS (Fig.~\ref{fig:ColorPlot}). For both H$\alpha$ and H$\beta$+[OIII] scenarios, a contamination of primeval extreme source cannot be ruled out.

In Tables~\ref{tab:nexp_Ha} and~\ref{tab:nexp_HbOIII}, we also report the fractions of H$\alpha$ and H$\beta$+[OIII] emission line galaxies compared to the total number of detections in MIDIS Deep and to the total number of detections with a positive color F444W-F560W>0. About 1\% of the detections in MIDIS are expected to be H$\alpha$ or H$\beta$+[OIII] emission line galaxies and among the ones with positive colors, almost half of them, 44\% for H$\alpha$ with EW$_0$=500 Å and 11\% (with the COSMOS-3D LF) and 47\% (with the FRESCO LF) for H$\beta$+[OIII] with EW$_0$=500 Å, are expected to be line emitters, consistent with the reddest sources in Fig.~\ref{fig:ColorPlot}. \cite{Matthee2024} reports that faint broad line AGNs from $M_{\rm UV}=-19$ to -21 are very rare, about 0.5\% and their abundance even decreases at fainter magnitudes. \cite{Tanaka2025_LRDphi} estimates that LRD at $M_{\rm UV}=-20$ reaches $2.6^{+5.9}_{-2.1}$ at $z\sim10$. It suggests two things: first, the selection of broad H$\alpha$ AGNs and V-shape LRDs with prominent a Balmer breaks are two different selections with different results, not probing the exact same group of galaxies. Second, broad line H$\alpha$ AGNs do not represent the main fraction of sources of reionization, but on the other hand, LRDs might play a significant role at very high redshifts. The MIRI wavelength coverage and deep observations such as MIDIS are necessary to address whether broad line AGNs, H$\alpha$ and H$\beta$+[OIII] emission line galaxies, and LRDs are significant populations of galaxies at very high redshifts and in what fraction. Our results suggest that a deeper MIDIS survey would reveal more H$\alpha$ and H$\beta$+[OIII] excess galaxies, and that a wider MIDIS survey is needed to catch an LRD with a strong spectral Balmer break at very high redshifts.

\begin{table}[H]
\caption{ Expected number and fractions of H$\alpha$ line emitters in MIDIS for different line equivalent widths, from a mapping between line luminosities and observed MIRI flux in MIDIS and the FRESCO luminosity functions \citep{Covelo-Paz2025_HaLF_fresco}. }
\label{tab:nexp_Ha}
\centering
\begin{tabular}{ c c c c }
\hline \hline
EW$_0$ [Å] & N(H$\alpha$) & f [\%] & f$_{\rm F444W-F560W>0}$ [\%] \\ \hline
500 & 13  & 1 & 44   \\
1000 & 9 & 1 & 31 \\
2000 & 7  & <1 & 23 \\
4000 & 6  & <1 & 19 \\
6000 & 5 & <1 & 17 \\

\hline
\end{tabular}
\end{table}

\begin{table}[H]
\caption{Expected number and fractions of H$\beta$+[OIII] line emitters in MIDIS for different line equivalent widths, from a mapping between line luminosities and observed MIRI flux in MIDIS and the FRESCO luminosity functions \citep{,Meyer2024_OIIILF_fresco}. }
\label{tab:nexp_HbOIII}
\centering
\begin{tabular}{ c c c c }
\hline \hline
EW$_0$ [Å] & N(H$\beta$+[OIII]) & f [\%] & f$_{\rm F444W-F560W>0}$ [\%]   \\ \hline
500  & 14 & 1 & 47  \\
1000 & 8 & 1 & 27  \\
2000 & 5 & <1 & 18  \\
4000 & 4 & <1 & 13  \\
6000 & 3 & <1 & 12  \\

\hline
\end{tabular}
\end{table}

\begin{table}[H]
\caption{Expected number and fractions of H$\beta$+[OIII] line emitters in MIDIS for different line equivalent widths, from a mapping between line luminosities and observed MIRI flux in MIDIS and the COSMOS-3D luminosity functions \citep{Meyer2025_OIIICOSMOS}. }
\label{tab:nexp_HbOIII_cosmos}
\centering
\begin{tabular}{ c c c c }
\hline \hline
EW$_0$ [Å] & N(H$\beta$+[OIII]) & f [\%] & f$_{\rm F444W-F560W>0}$ [\%]   \\ \hline

500 & 3  & <1  & 11  \\
1000 & 2  & <1  & 7  \\
2000 & 1  & <1  & 5  \\
4000 & 1  & <1  & 4  \\
6000 & 1  & <1  & 4  \\

\hline
\end{tabular}
\end{table}

\section{Summary}
\label{sec:Summary}

We present an in-depth analysis of the MIri Deep Imaging Survey (MIDIS), the deepest 5.6\,$\mu$m MIRI imaging of the Hubble Ultra Deep Field, including a search for MIRI-red galaxies requiring strong spectral breaks between NIRCam/F444W and MIRI/F560W. We perform a detailed characterization of the MIRI background and demonstrate that standard pipeline uncertainties overestimate the signal-to-noise ratio of faint sources by a factor of $\sim3$. With an empirical approach, we estimate the purity and completeness of the MIDIS catalog and find that a 5$\sigma$ detection at 28.75 mag has a purity of 92\% and a completeness of 54\%.

We identify seven MIRI-red galaxy candidates with $5.00 < \mathrm{S/N}_{\rm MIRI} < 33.17$, including one F115W dropout consistent with a high-redshift galaxy candidate, \textit{Virgil}, one spurious source, and one MIRI-only candidate undetected in NIRCam, which is likely spurious due to PSF artifacts. In total, we report five robust MIRI-red galaxy candidates. All spectroscopically confirmed red galaxies in MIDIS are extreme emission-line galaxies dominated by H$\alpha$ or H$\beta$+[OIII] with rest-frame equivalent widths of a few hundred angstroms. SEDs fits of the MIRI-red galaxy candidates show preferred solutions with a template of a black hole star galaxy, and including deeper longer wavelength MIRI observation allows for a better constrain of emission lines in F560W. 

Given the proximity of the F444W and F560W filters, we find that potential MIRI-only detections would require unrealistic extreme emission-line equivalent widths (EW$_0({\rm H\alpha})\sim6000$ Å or EW$_0({\rm H\beta+[OIII]})\sim4000$ Å) unless it invokes an extreme ionizing source such as a Pop III star or a DCBH, or very strong Balmer breaks, consistent with high-redshift analogs of spectroscopically observed Little Red Dots. 

MIRI-red galaxies are consistent with strong emission-line equivalent widths (EW$_0({\rm H\alpha})\sim 750$ Å or EW$_0({\rm H\beta+[OIII]})\sim600$ Å). We find fewer MIRI-red detections than expected from extrapolations of the FRESCO spectroscopic luminosity functions, in agreement with previous deep searches in the footprint of MIDIS. Accounting for AGN contamination in the estimation of expected H$\alpha$ emitters and for cosmic variance in the [OIII] luminosity function solves the discrepancy. These results highlight both the scientific potential and current limitations of the \textit{JWST} capabilities through ultra-deep MIRI imaging for identifying the faintest and most extreme galaxies at cosmic dawn.

\begin{acknowledgements}

IJ and TRG are grateful for support from the Carlsberg Foundation via grant No.~CF20-0534. The Cosmic Dawn Center (DAWN) is funded by the Danish National Research Foundation under grant No. 140. LC, JAM, CPJ acknowledges support from grant PID2021-127718NB-I00 and PID2024-158856NA-I00, PID2024-159902NA-I00 funded by Spanish Ministerio de Ciencia e Innovación MCIN/AEI/10.13039/501100011033 and by “ERDF A way of making Europe. The project that gave rise to these results received the support of a fellowship from the “la Caixa” Foundation (ID 100010434). The fellowship code is LCF/BQ/PR24/12050015. LC acknowledges support from grants PID2022-139567NB-I00 and PIB2021-127718NB-I00 funded by the Spanish Ministry of Science and Innovation/State Agency of Research  MCIN/AEI/10.13039/501100011033 and by “ERDF A way of making Europe”. This work was supported by research grants (VIL16599,VIL54489) from VILLUM FONDEN. L.A.B. acknowledges support from the Dutch Research Council (NWO) under grant VI.Veni.242.055 (\url{https://doi.org/10.61686/LAJVP77714}). JPP and TVT acknowledge financial support from the UK Science and Technology Facilities Council, and the UK Space Agency. For the purpose of open access, the authors have applied a Creative Commons Attribution (CC BY) licence to the Author Accepted Manuscript version arising from this submission. The following National and International Funding Agencies funded and supported the MIRI development: NASA; ESA; Belgian Science Policy Office (BELSPO); Centre Nationale d’Etudes Spatiales (CNES); Danish National Space Centre; Deutsches Zentrum fur Luftund Raumfahrt (DLR); Enterprise Ireland; Ministerio De Economiá y Competividad; Netherlands Research School for Astronomy (NOVA); Netherlands Organisation for Scientific Research (NWO); Science and Technology Facilities Council; Swiss Space Office; Swedish National Space Agency; and UK Space Agency.
MIRI drew on the scientific and technical expertise of the following organizations: Ames Research Center, USA; Airbus Defence and Space, UK; CEAIrfu, Saclay, France; Centre Spatial de Liège, Belgium; Consejo Superior de Investigaciones Cientficas, Spain; Carl Zeiss Optronics, Germany; Chalmers University of Technology, Sweden; Danish Space Research Institute, Denmark; Dublin Institute for Advanced Studies, Ireland; European Space Agency, Netherlands; ETCA, Belgium; ETH Zurich, Switzerland; Goddard Space Flight Center, USA; Institute d’Astrophysique Spatiale, France; Instituto Nacional de Tecnica Aeroespacial, Spain; Institute for Astronomy, Edinburgh, UK; Jet Propulsion Laboratory, USA; Laboratoire d’Astrophysique de Marseille (LAM), France; Leiden University, Netherlands; Lockheed Advanced Technology Center (USA); NOVA Opt-IR group at Dwingeloo, Netherlands; Northrop Grumman, USA; Max Planck Institut für Astronomie (MPIA), Heidelberg, Germany; Laboratoire d’Etudes Spatiales et d’Instrumentation en Astrophysique (LESIA), France; Paul Scherrer Institut, Switzerland; Raytheon Vision Systems, USA; RUAG Aerospace, Switzerland; Rutherford Appleton Laboratory (RAL Space), UK; Space Telescope Science Institute, USA; Toegepast- Natuurwetenschappelijk Onder19zoek (TNOTPD), Netherlands; UK Astronomy Technology Centre, UK; University College London, UK; University of Amsterdam, Netherlands; University of Arizona, USA; University of Cardiff , UK; University of Cologne, Germany; University of Ghent; University of Groningen, Netherlands; University of Leicester, UK; University of Leuven, Belgium; University of Stockholm, Sweden; Utah State Uni, USA. We thank A. Weibel for kindly sharing with us the BH-$\star$ template and suggesting the SED fitting procedure.

\end{acknowledgements}

\hypertarget{orcids}{\section*{ORCIDs}}
Iris Jermann \orcid{0000-0002-2624-1641}\\
Gabe Brammer \orcid{0000-0003-2680-005X} \\ 
Steven Gillman \orcid{0000-0001-9885-4589} \\
Thomas R. Greve \orcid{0000-0002-2554-1837]}\\
Leindert Boogaard \orcid{0000-0002-3952-8588}\\
Pablo G. P\'erez-Gonz\'alez \orcid{0000-0003-4528-5639} \\
Jens Melinder \orcid{0000-0003-0470-8754}\\
Romain A. Meyer \orcid{0000-0001-5492-4522}\\
Pierluigi Rinaldi \orcid{0000-0002-5104-8245} \\
Luis Colina \orcid{0000-0002-9090-4227}\\
G{\"o}ran {\"O}stlin \orcid{0000-0002-3005-1349}\\
Gillian Wright \orcid{0000-0001-7416-7936}\\
Javier Álvarez-Márquez \orcid{0000-0002-7093-1877} \\
Arjan Bik \orcid{0000-0001-8068-0891}\\
Karina I. Caputi \orcid{0000-0001-8183-1460} \\
Alejandro Crespo G\'omez \orcid{0000-0003-2119-277X}\\
Luca Costantin \orcid{0000-0001-6820-0015}\\
Jens Hjorth \orcid{0000-0002-4571-2306}\\
Edoardo Iani\orcid{0000-0001-8386-3546}\\
Sarah Kendrew\orcid{0000-0002-7612-0469}\\
\'Alvaro Labiano \orcid{0000-0002-0690-8824}\\
Danial Langeroodi\orcid{0000-0001-5710-8395} \\
Florian Peissker \orcid{0000-0002-9850-2708} \\
Carlota Prieto-Jiménez \orcid{0009-0005-4109-161X}\\
John P. Pye \orcid{0000-0002-0932-4330}\\
Tuomo Tikkanen \orcid{0009-0003-6128-2347}\\
Fabian Walter \orcid{0000-0003-4793-7880} \\
Paul P.~van der Werf \orcid{0000-0001-5434-5942]} \\
Thomas Henning \orcid{0000-0002-1493-300X} \\
Marko Shuntov \orcid{0000-0002-7087-0701}

\bibliographystyle{aa}
\bibliography{ReferenceFileClean}

\newpage
\begin{appendix}

\section{The other reduction versions}
\label{sec:AppendixA_MIDIS_versions}

In this appendix, we report the depths of the images with the two different reductions on the two pixel scales in Table~\ref{table:Appendix_MIDISareadepth}. The values are consistent with the depth reported in \citet{Ostlin2025} (see Sects.~\ref{sec:MIDIS} and~\ref{sec:Depths}). The earlier reductions underestimate uncertainties more strongly, suggesting improved noise propagation in the newer JWST pipeline version. A smaller pixel scale implies errors underestimated by a bigger factor.

\begin{table}[H]
\caption{ The areas in arcmin$^2$ and the 5$\sigma$ aperture corrected point-source depths of the 5.6\,$\mu$m MIDIS Deep observations. 
}
\label{table:Appendix_MIDISareadepth}
\centering
\begin{tabular}{c|c|c}
\hline \hline

This paper & Area [arcmin$^2$] & depth [5$\sigma$]\\ \hline

MIDIS deep v.1.4 0\farcs{04} & 2.4 & 28.69   \\
MIDIS deep v.2.2.8 0\farcs{04} & 2.5 &  28.75   \\
MIDIS deep v.1.4 0\farcs{06} & 2.4 &  28.58   \\
MIDIS deep v.2.2.8 0\farcs{06} & 2.6 &  28.68   \\ \hline \hline 

\citet{Ostlin2025} & Area & depth \\ \hline
A & 1.12 & 28.65 \\
B & 1.64 & 28.48 \\
\hline
\end{tabular}
\end{table}

\begin{figure}[h] 
   \centering
  \includegraphics[width=\hsize]{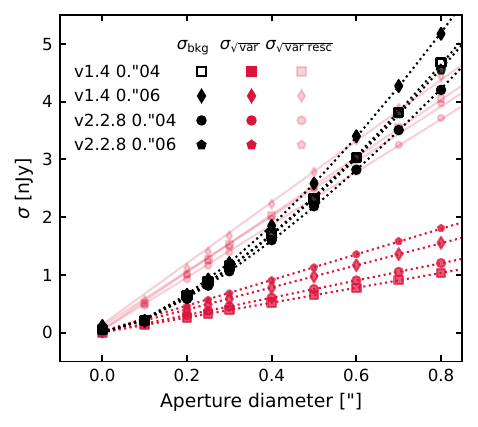}
   \caption{ Background noise level as a function of aperture diameter of MIDIS Deep v1.4 and v2.2.8, and in two pixel scales. The black markers indicate the background noise level derived from the scatter in empty background apertures. The red markers denote the average error derived from the sum in quadrature of the pixels in the associated variance map to the MIDIS observations. The shaded red markers denote the average error derived after rescaling the variance maps. The background noise levels ($\sigma_{\rm bkg}$) are used as the 1$\sigma$ error for photometric measurements in given aperture sizes, as the average errors ($\sigma_{\rm var}$) are underestimated.  }

\label{fig:Appendix_NoiseStudyFigAll}
\end{figure}

\begin{table*}[h]
\caption{ Best-fit parameters to Eq.~\ref{eqn:PowerLaw} and Eq.~\ref{eqn:Linear}) to the noise measured in empty background apertures. $\eta$ are the rescaling factors to apply to the associated total variance map ($\sqrt{\eta}$ for the error map) to derive accurate photometric uncertainties from the MIDIS images. }
\label{table:Appendix_NoiseResultsAll}
\centering
\begin{tabular}{  c| l|ll|ll}
\hline \hline

 &   & \multicolumn{2}{c|}{MIDIS v1.4} & \multicolumn{2}{c}{MIDIS v2.2.8} \\ 
 &   &  0\farcs{04}         &   0\farcs06        &  0\farcs{04}    &  0\farcs06 \\ \hline
(1) $\sigma_{\rm background} = \sigma_{\rm background}^{\rm per~pixel}\alpha D^\beta$ & $\sigma_{\rm per pixel}$ & 0.053  &  0.114 &  0.049  &  0.107   \\
& $\alpha$ &  123.1 $\pm$ 0.7 &  63.8 $\pm$ 0.3 &  117.7 $\pm$ 0.4 &  58.7 $\pm$ 0.2          \\
& $\beta$ &  1.48 $\pm$ 0.01 &  1.49 $\pm$ 0.01 &  1.41 $\pm$ 0.01 &  1.44 $\pm$ 0.01     \\  \hline
& $\mu_{\rm per pixel}$ &  0.003  &  0.018  &  0.005  &  0.023      \\
(2) $\sigma_{\rm \sqrt{VAR}} = \sigma_{\rm \sqrt{VAR}}^{\rm per~pixel} + a D$  & $a$ &   1.29 $\pm$ 0.01 &  1.91 $\pm$ 0.01 &  1.50 $\pm$ 0.01 &  2.22 $\pm$ 0.01    \\
& $a_{\rm background}$ &    5.0 $\pm$ 0.3 &  5.5 $\pm$ 0.3 &  4.6 $\pm$ 0.2 &  4.9 $\pm$ 0.3   \\ \hline

(3) $\eta=\frac{a}{a_{\rm background}}$ & $\eta$ & 3.9 $\pm$ 0.2 &  2.9 $\pm$ 0.2 &  3.1 $\pm$ 0.2 &  2.2 $\pm$ 0.1     \\

\hline 
\end{tabular}
\end{table*}

\section{ Purity and number of candidates with and without non-masked regions }
\label{sec:AppendixB_no_mask}

In this appendix, we provide the data points used to compute the purity and completeness of the MIDIS survey in Tables~\ref{tab:Appendix_PurityData} and~\ref{tab:Appendix_ComplData}, as presented in Sect.~\ref{sec:MIDISdeepPurityCompleteness} and shown in Fig.~\ref{fig:MIRIpurity}.

\begin{table}[H]
\caption{Measured purities data points and binomial uncertainties shown in Fig.~\ref{fig:MIRIpurity}}
\label{tab:Appendix_PurityData}
\begin{tabular}{ c c }
\hline \hline
MIRI [5.6\,mag] & Purity [\%]  \\ \hline
 29.98 & 33.33$\pm$ 27.22   \\
 29.48 & 67.14$\pm$ 5.61   \\
 29.13 & 86.49$\pm$ 5.00   \\
 28.85 & 91.74$\pm$ 5.00   \\
 28.66 & 89.80$\pm$ 5.00   \\
 28.47 & 96.55$\pm$ 5.00   \\
 28.32 & 96.49$\pm$ 5.00   \\
 28.17 & 100.00$\pm$ 5.00   \\
 28.06 & 100.00$\pm$ 5.00   \\
 27.95 & 94.44$\pm$ 5.00   \\
 27.85 & 96.77$\pm$ 5.00   \\
 27.75 & 100.00$\pm$ 5.00   \\
 27.67 & 95.45$\pm$ 5.00   \\
 27.60 & 95.00$\pm$ 5.00   \\
 27.52 & 100.00$\pm$ 5.00   \\
 27.45 & 100.00$\pm$ 5.00   \\
 27.39 & 100.00$\pm$ 5.00   \\
 27.32 & 100.00$\pm$ 5.00   \\
 27.27 & 100.00$\pm$ 5.00   \\
 27.22 & 100.00$\pm$ 5.00   \\
 27.16 & 100.00$\pm$ 5.00   \\
 27.13 & 100.00$\pm$ 5.00   \\
 27.08 & 100.00$\pm$ 5.00   \\
 27.03 & 75.00$\pm$ 21.65   \\
 26.98 & 100.00$\pm$ 5.00   \\
 26.94 & 100.00$\pm$ 5.00   \\
 26.90 & 100.00$\pm$ 5.00   \\
 26.86 & 100.00$\pm$ 5.00   \\
 26.82 & 100.00$\pm$ 5.00   \\
\hline
\end{tabular}
\end{table}

\begin{table}[H]
\caption{Measured completeness data points and binomial uncertainties shown in Fig.~\ref{fig:MIRIpurity}}
\label{tab:Appendix_ComplData}
\begin{tabular}{ c c }
\hline \hline
MIRI [5.6\,mag] & Completeness [\%]  \\ \hline
 31.74 & 0.00$\pm$ 1.00   \\
 30.18 & 0.00$\pm$ 1.00   \\
 30.06 & 0.00$\pm$ 1.00   \\
 29.82 & 0.00$\pm$ 1.00   \\
 29.70 & 100.00$\pm$ 1.00   \\
 29.58 & 33.33$\pm$ 27.22   \\
 29.46 & 0.00$\pm$ 1.00   \\
 29.34 & 50.00$\pm$ 35.36   \\
 29.22 & 0.00$\pm$ 1.00   \\
 29.10 & 33.33$\pm$ 27.22   \\
 28.98 & 27.27$\pm$ 13.43   \\
 28.86 & 50.00$\pm$ 15.81   \\
 28.74 & 60.00$\pm$ 12.65   \\
 28.62 & 35.00$\pm$ 10.67   \\
 28.50 & 63.64$\pm$ 10.26   \\
 28.38 & 70.37$\pm$ 8.79   \\
 28.26 & 71.79$\pm$ 7.21   \\
 28.14 & 86.49$\pm$ 5.62   \\
 28.02 & 91.67$\pm$ 3.99   \\
 27.90 & 96.43$\pm$ 2.48   \\
 27.78 & 98.18$\pm$ 1.80   \\
 27.66 & 96.23$\pm$ 2.62   \\
 27.54 & 100.00$\pm$ 1.00   \\
 27.42 & 100.00$\pm$ 1.00   \\
 27.30 & 100.00$\pm$ 1.00   \\
 27.18 & 100.00$\pm$ 1.00   \\
 27.06 & 100.00$\pm$ 1.00   \\
\hline
\end{tabular}
\end{table}

\section{NIRCam and MIRI broadband cutouts of the analysis sample}
\label{sec:AppendixC1_Cutouts}

\begin{figure*}[h!] 
\centering
   \includegraphics[width=1.0\hsize]{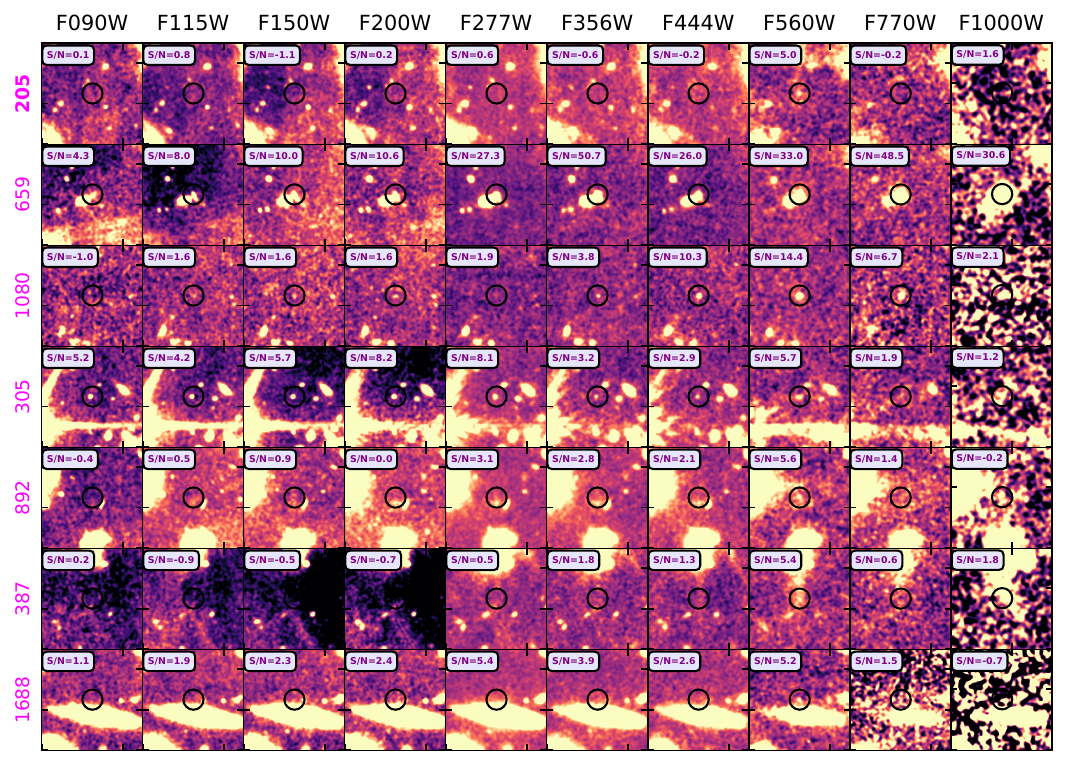}
      \caption{NIRCam broadbands and MIRI F560W, F770W, and F1000W 5"x5" cutouts of the MIRI-red galaxy candidates (Eq.~\ref{eqn:MIRIselection}) in MIDIS Deep. }
         \label{fig:MIRIonly_cutouts}
\end{figure*}

In this appendix, we show in Fig.~\ref{fig:MIRIonly_cutouts} the cutouts of MIRI-red galaxy candidates along with the two reddest spectroscopically confirmed sources in MIDIS with NIRSpec/PRISM.

\section{Expected number of H$\alpha$-excess, H$\beta$+[OIII]-excess, and LRDs}
\label{sec:AppendixC_ExpNumber}

In this appendix, we describe the methodology to estimate the expected number of galaxies in the MIDIS survey for different galaxy populations. First, we estimate the expected number of emission-line galaxies with a flux excess in the  MIRI/F560W band due to an H$\alpha$ or H$\beta$+[OIII] emission line. This calculation is based on the results of the FRESCO survey \citep{Oesch2023_FRESCO}. We adopt the Schechter-function best-fit parameters of the spectroscopic luminosity functions for H$\alpha$ emitters at $z\sim4-6.5$ \citep{Covelo-Paz2025_HaLF_fresco} and for H$\beta$+[OIII] emitters at $z\sim6.8-9.0$ \citep{Meyer2024_OIIILF_fresco}. We assume no evolution of the luminosity functions when extrapolation from the GRISM/F444W to our estimates in F560W for the following reasons: our estimate is based on broadband photometry, which is naturally limited in its accuracy; the FRESCO luminosity functions show a weak redshift evolution over the redshift range covered in F444W; at the faint-end, the uncertainties are large and the statistics are sparse; and there are degeneracies among the Schechter best-fit parameters. Next, we compute the expected number of little red dots given the survey area and depth of MIDIS and the theoretical LRD number density model from \cite{Inayoshi2025}. \\

The main challenge in estimating the expected number of H$\alpha$ or H$\beta$+[OIII] emitters detected in F560W is defining a mapping between the intrinsic emission-line luminosity and the observed flux density in F560W. Specifically, we convert line luminosities expressed in units of erg s$^{-1}$ into observed flux densities:

\begin{equation}
    \label{eqn:Mapping}
L_{\rm line}(z,{\rm EW}_0)\,[{\rm erg\,s^{-1}}]\longrightarrow L_{\rm line}(z,{\rm EW}_0)\,[{\rm erg\,s^{-1}\,cm^{-2}\,Hz^{-1}}]
\end{equation}

 To perform this conversion, we construct SED templates for emission-line galaxies with rest-frame equivalent widths, EW$_0$=500, 1000, 2000, 4000, and 6000 Å. Each template consists of the sum of a continuum and a single Gaussian emission line. We assume a flat continuum in $F_\nu$, parametrized as:
\begin{equation}
    \label{eqn:MappingContinuum}
F_\nu\,=\,\left( \frac{\lambda}{\lambda_0} \right)^\alpha
\end{equation}

where $\lambda_0$ is the rest-frame line center and $\alpha=-2$. The Gaussian emission line is set to a FWHM of 0.001\,$\mu$m, corresponding to approximately 1\% of the F560W bandwidth. The templates are defined on a rest-frame wavelength grid from 300 Å to 70000 Å. We use the \textsc{eazy.template} package and filter curves to calculate the observed flux densities in F560W for different redshifts and line luminosities. 

For all both H$\alpha$ and H$\beta$+[OIII], we sample a redshift grid using the wavelength enclosing 10\% to the wavelength enclosing 90\% of the total flux in the F560W bandpass, and the rest-frame line center wavelengths ($z_{\rm min}$ and $z_{\rm max}$ in Table~\ref{tab:RequMIRIonly}). On these redshift grids, we then evaluate the Schechter function as a function of flux in MIRI, for all templates with the different EW$_0$, with the best-fit values from \cite{Covelo-Paz2025_HaLF_fresco} and \cite{Meyer2024_OIIILF_fresco}:

\begin{equation}
    \label{eqn:logphi}
    \phi(\log F_{\rm MIRI}|z,{\rm EW}_0)\,[{\rm N\,dex^{-1}\,Mpc^{-3}}]
\end{equation}

where $F_{\rm MIRI}$ is the flux density in units of erg s$^{-1}$ cm$^{-2}$ Hz$^{-1}$. For each value of EW$_0$, we interpolate the $\phi(\log F_{\rm MIRI}|z,{\rm EW}_0)$ onto a common $\log F_{\rm MIRI}$ grid and integrate over redshift, from $z_{\rm min}$ to $z_{\rm max}$ (see Table~\ref{tab:RequMIRIonly}), to obtain the differential number counts:
 
\begin{equation}
    \label{eqn:LF}
    \frac{dN}{d\log F}=\phi(F_{\rm MIRI}|EW_0) = \Omega \int^{z_{\rm max}}_{z_{\rm min}}\phi(F_{\rm MIRI}|z,EW_0) \frac{dV}{dzd\Omega} dz
\end{equation}

where $\Omega$ is the survey area and $\frac{dV}{dzd\Omega}$ is the differential comoving volume element, computed using \textsc{astropy}. The total expected number of  H$\alpha$ and H$\beta$+[OIII] emitters in MIDIS is then obtained by integrating the differential number counts over the F560W grid and applying the survey completeness correction (see Sect.~\ref{sec:MIDISdeepPurityCompleteness}, Eq.~\ref{eqn:erf_compl}, and Fig.~\ref{fig:MIRIpurity}):

\begin{equation}
    \label{eqn:LF2}
    N_{\rm tot} = \int^{\log F_{\rm max}}_{\log F_{\rm min}} C(F)\phi(F_{\rm MIRI}|EW_0) d\log F
\end{equation}

Finally, we compute the cumulative expected number of emission-line galaxies as a function of F560W flux density, which we compare with the observed MIDIS number counts (see Fig.~\ref{fig:NumberCounts} and Sect.~\ref{sec:NumberDensity}).

\begin{equation}
    \label{eqn:LF3}
    N(>F) = \int^{\log F_{\rm max}}_{\log F} C(F^\prime)\phi(F^\prime_{\rm MIRI}|EW_0) d\log F^\prime
\end{equation}

For the expected number of LRDs in MIDIS, we compute the integral in Eq.~\ref{eqn:LF2}, replacing the emission-line luminosity function with the LRD comoving number density model presented in \cite{Inayoshi2025}, Eqn.~2 and~3, and adopting the redshift range over which a Balmer break falls within the F560W bandpass (Table~\ref{tab:RequMIRIonly} and Sect.~\ref{sec:OnTheNature}).

\begin{figure}[h] 
   \centering
  \includegraphics[width=\hsize]{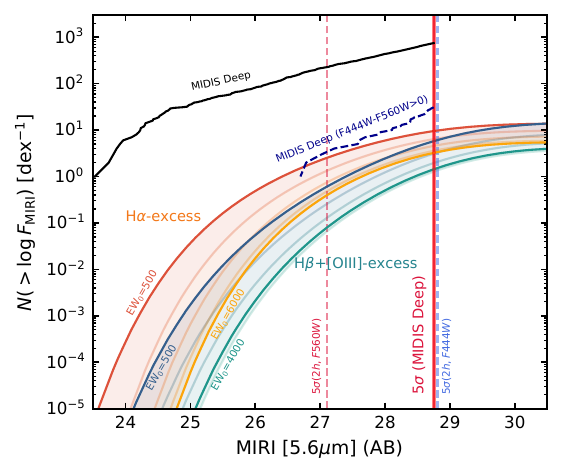}
   \caption{ Same as Fig.~\ref{fig:NumberCounts}
   MIDIS Deep survey number counts for detections with F560W magnitudes $>23.5$ mag and aperture-corrected fluxes measured in apertures of diameter D\,=\,0\farcs{4}. These are compared to predicted number counts for H$\alpha$ and H$\beta$+[OIII] emitters derived from observed line luminosity functions from FRESCO \citep{Covelo-Paz2025_HaLF_fresco,Meyer2024_OIIILF_fresco} for different values of emission-line rest-frame equivalent width.  }

\label{fig:NumberCountsAppendix}
\end{figure}

\section{Spectral Energy Fitting with F560W only}
\label{sec:AppendixD_SED}

\begin{figure}[h!] 
\centering
   \includegraphics[width=1.0\hsize]{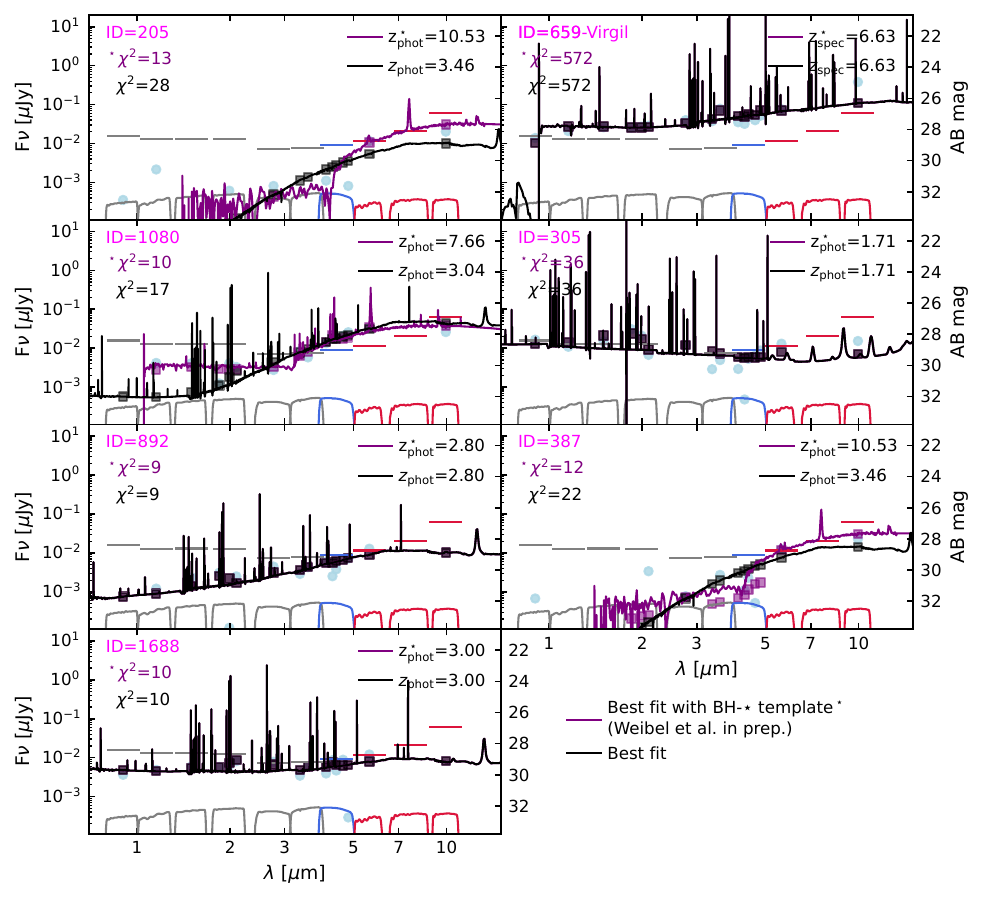}
      \caption{ Same as Fig.~\ref{fig:MIRIonly_red_SED} with MIDIS data only (no F770W and $\sim$10h in F1000W). }
         \label{fig:MIRIonly_red_SED_noF700WshallowF1000W}
\end{figure}

In this appendix, we show the best fits with \textsc{EAZY} without the deep F770W and F1000W images and the photometric redshift probability distributions to the fits in Figs.~\ref{fig:MIRIonly_red_SED} and~\ref{fig:MIRIonly_red_SED_noF700WshallowF1000W}.

\begin{figure}[h!] 
\centering
   \includegraphics[width=1.0\hsize]{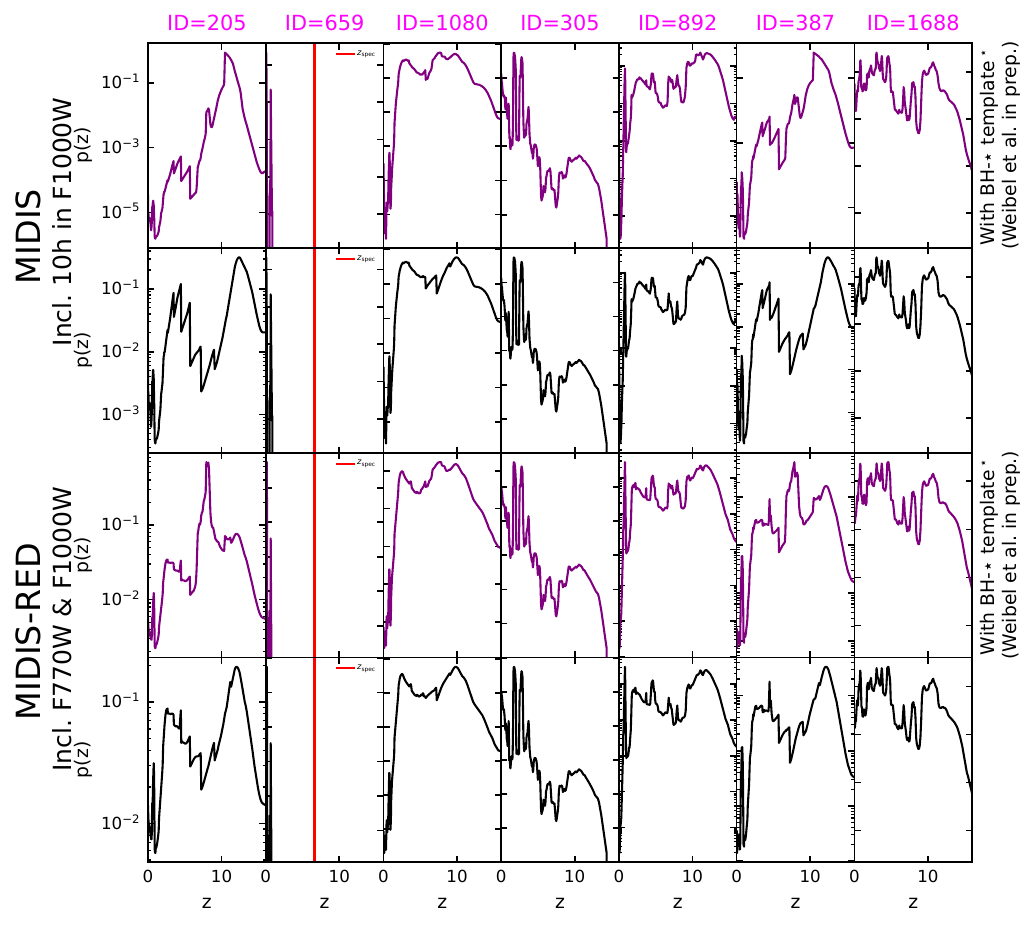}
      \caption{Photometric redshift probability distributions to fits in Figs.~\ref{fig:MIRIonly_red_SED} and~\ref{fig:MIRIonly_red_SED_noF700WshallowF1000W}.}
         \label{fig:pofz}
\end{figure}

\end{appendix}

\end{document}